\documentclass[useAMS,usenatbib, usegraphicx]{mn2e} 
\usepackage{aasmacros, amsmath}   
\newcommand{\gps}{\ensuremath{g_{\rm P1}}}
\newcommand{\rps}{\ensuremath{r_{\rm P1}}}
\newcommand{\ips}{\ensuremath{i_{\rm P1}}}
\newcommand{\zps}{\ensuremath{z_{\rm P1}}}
\newcommand{\yps}{\ensuremath{y_{\rm P1}}}

\newcommand{\grizy}{\gps, \rps, \ips, \zps and \yps}

\newcommand{\plotPath}{figures}
\newcommand{\wthFolder}{hamilton_plots}

\title[PS1: Testing Galaxy Clustering with SAS2]{Pan-STARRS1: Galaxy Clustering in the Small Area Survey 2}
\author[Daniel J. Farrow ]{Daniel J. Farrow$^{1}$\thanks{E-mail:d.j.farrow@durham.ac.uk}, 
Shaun Cole$^{1}$, N. Metcalfe$^{2}$, P. W. Draper$^{1,2}$, Peder Norberg$^{1}$, \and 
S{\'e}bastien Foucaud$^{3, 4}$, W. S. Burgett$^{5}$, K. C. Chambers$^{5}$, N. Kaiser$^{5}$, R. P. Kudritzki$^{5}$, \and
E. A. Magnier$^{5}$, P. A. Price$^{6}$, J. L. Tonry$^{5}$, C. Waters$^{5}$\\
$^{1}$Institute for Computational Cosmology, Department of Physics, Durham University, South Road, DH1 3LE, U.K.\\
$^{2}$Department of Physics, Durham University, South Road, DH1 3LE, U.K.\\
$^{3}$National Taiwan Normal Univeristy, Department of Earth Sciences, 88 Tingzhou Road, Sec. 4, Wenshan district, Taipei 11677, Taiwan\\
$^{4}$Institute of Astronomy and Astrophysics, Academia Sinica, P.O. Box 23-141, Taipei 10617, Taiwan\\
$^{5}$Institute for Astronomy, University of Hawaii, 2680 Woodlawn Drive, Honolulu HI 96822, U.S.A.\\
$^{6}$Department of Astrophysical Sciences, Princeton University, Princeton, NJ 08544, U.S.A.\\
}
\begin{document}

\date{}

\pagerange{\pageref{firstpage}--\pageref{lastpage}} \pubyear{2013}

\maketitle

\label{firstpage}

\begin{abstract}

The Pan-STARRS1 survey is currently obtaining imaging in 5 bands (\grizy) for the $3\pi$ steradian survey, one of the largest optical surveys ever conducted. The finished survey will have spatially varying depth, due to the survey strategy. This paper presents a method to correct galaxy number counts and galaxy clustering for this potential systematic based on a simplified signal to noise measurement. A star and galaxy separation method calibrated using realistic synthetic images is also presented, along with an approach to mask bright stars. By using our techniques on a $~69$ sq. degree region of science verification data this paper shows PS1 measurements of the two point angular correlation function as a function of apparent magnitude agree with measurements from deeper, smaller surveys. Clustering measurements appear reliable down to a magnitude limit of $\rps<22.5$. Additionally, stellar contamination and false detection issues are discussed and quantified. This work is the second of two papers which pave the way for the exploitation of the full $3\pi$ survey for studies of large scale structure.   

\end{abstract}


\section{Introduction}

Pan-STARRS1 (PS1) is a 1.8m telescope on Haleakala, Maui \citep{hodapp}. Its unique selling point is its high \emph{etendue}, the product of its collecting area and field of view, which allows it to survey large areas of sky quickly \citep{kaiser02}. To fully utilise this it has a huge camera (GPC1), with 1.4 Gpixels and a 3.3 degree field of view \citep{tonry08}. It was designed as a prototype of PS4, an array of four identical telescopes scanning the whole sky in relatively short intervals for potentially threatening Near Earth Objects (NEOs) \citep{kaiser02}. The multiepoch nature of PS1 observations is not only good for the detection of moving and transient objects but also provides the redundancy necessary for highly accurate zero point calibration \citep{eddie2, magnier13}, which is important for large scale structure analysis.  

As well as the main goal of detecting NEOs, PS1 has always been envisaged to meet a wide variety of science goals, including comets, extra-solar planets, supernovae, AGNs and large scale structure. PS1 does not have a spectrograph but photometric redshifts will be available from a dedicated pipeline \citep{saglia12}. As of July 2013, PS1 has been successful in detecting many new solar system objects\footnote[2]{http://www.minorplanetcenter.org/iau/mpc.html}, as well as supernovae \citep[e.g.][]{smartt}, variable AGN \citep[e.g.][]{ward} and satellite galaxies around Andromeda \citep{martin2013}. It has also been successfully used as a source of optical data for other surveys to measure the clustering of Extremely Red Galaxies (Kim et al. in preparation). We now extend this success to large scale structure using PS1 data alone with static objects, where individual exposures are co-added to gain greater depth.  

The finished PS1 survey will have two major co-added data products. The 3$\pi$ survey with 31,500 square degrees of imaging and ten deeper 8.5 square degree fields known as the ``Medium Deeps'', both in the PS1 bands of \grizy. The 3$\pi$ survey will be deeper and have a larger area than its predecessors, and it will also benefit from, \yps, a near infrared band. For more details on the 3$\pi$ survey please refer to Chambers et al. (in preparation). 

In this work we lay the foundations of exploiting the 3$\pi$ survey for large scale structure by demonstrating how galaxy number counts and the angular two point galaxy correlation function, $w(\theta)$, can be reliably measured. Namely we tackle, from a large scale structure stand-point, issues of star/galaxy separation, false positives, depth, angular masks and how completeness varies with sky position. We will refer to the fraction of objects detected as a function of magnitude as the ``detection efficiency'' throughout. As this paper is mainly a proof of concept, we concentrate mostly on the \rps-band. This is the second of two papers assessing PS1 viability for large scale structure studies, we will refer to our first paper, \citet{metcalfe2013}, as Paper I hereafter. 

This paper is organised as follows. In Section 2 we introduce the data sets we are using from PS1 along with the SDSS comparison samples. In Section 3 we present the angular masks and in Section 4 we create synthetic images and use them to define star and galaxy separators. Section 5 introduces our method of dealing with spatially varying depth. In Section 6 we present our measurements of clustering and number counts along with careful tests of how systematic errors and our corrections affect them. In Section 7 we discuss implications of our work to the scientific exploitation of the finished 3$\pi$ survey.

\section[Data]{The Data}
\subsection{The PS1 small area survey 2}\label{sec:data}
The Small Area Survey 2 (SAS2) is a subset of the $3\pi$ survey roughly covering the region of $327.5<\alpha\rm{(deg.)}<338.5$ and $-5.5<\delta\rm{(deg.)}<5.5$. It is designed to be representative of the finished $3\pi$ survey. A large number of individual exposures were taken, co-added and mosaiced to form around 69 square degrees of imaging. It has a median PSF FWHM of $0.94\arcsec$, which has an rms scatter of less than 0.05\arcsec across the field (Paper I). PS1 has a raw pixel scale of $0.256\arcsec$ before ``warping'' (see later in this section) and $0.25\arcsec$ after. A careful study of the depth of this data set can be found in Paper I, which reports that 50\% of stars are recovered at magnitudes in \gps,\rps,\ips,\zps and \yps of 23.4, 23.4, 23.2, 22.4 and 21.3 respectively.  All magnitudes in this paper are measured in the AB system.

Different sub-areas of the finished SAS2 stacked data have different numbers of input exposures. This is down to the observing strategy, which means exposures in a stack are not always coincident with each other. Additionally around 25\% of individual exposures are masked (Paper I), which is mainly due to gaps between CCD chips as well as defective CCD cells and other regions. The decision to build up stacks using multiple, rotated and non-coincident individual exposures was chosen in order to meet the needs of scientists interested in transient and moving objects, who require large area imaging over multiple epochs.  

\begin{figure}
\begin{center}
  \includegraphics[width=3.8in]{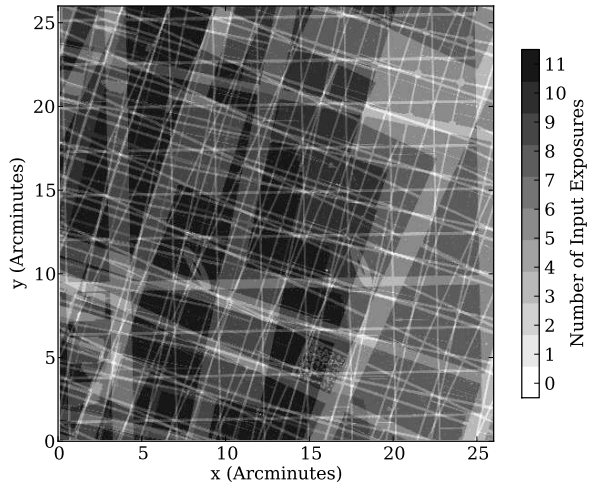}
  \caption{The coverage, i.e. the number of input exposures, of a typical 26 by 26 arcminute SAS2 stack skycell. Black areas correspond to 11 input exposures for that pixel, white corresponds to no input exposures (a blank pixel). The grid pattern arises from the gaps between CCD chips in individual exposures.}\label{fig:cov}
\end{center}	
\end{figure}

We will refer to the number of input exposures to a pixel as the ``coverage'' throughout this paper. To illustrate this Fig. \ref{fig:cov} gives the ``coverage map'', i.e. an image recording the number of exposures stacked for each pixel, in a $26'$ by $26'$ region. A typical SAS2 stacked image has an average coverage of around 8.9 exposures per pixel (Paper I), and this coverage has a standard deviation of around 3 exposures per pixel. In the stacks this gives rise to a spatially varying noise level. To track this PS1 produces ``variance maps'' which record the variance of the noise in each image pixel. This variance includes contributions from sources of astronomical noise including sky background, read noise and Poisson noise, and how they scale with the weighting of exposures in a stack. Naturally the spatially varying image noise leads to different depths in different positions on the sky (see Section \ref{sec:detectionEff}).

In addition to coverage maps and variance maps the PS1 Image Processing Pipeline (IPP) \citep{magnier2006} also produces image masks. These image masks track pixel quality and highlight pixels which have been flagged as suspicious (e.g. likely to be cosmic rays or image artifacts) by the pipeline. Image masks, variance maps, coverage maps and images are all supplied in approximately $26'$ by $26'$ units called ``skycells''. These skycells do not represent unique areas on the sky but overlap, and in these overlap regions pixels from different skycells are not necessarily the same, since decisions on which exposures to reject from a stack are made on a skycell by skycell basis.

It is also important to note that transforming the exposures from the CCD coordinates to the stack pixel coordinate system, a process known as ``warping'', introduces correlations between the image pixels on scales of less than around $1''$ (see Paper I). The image, $I$, the variance $V$ and the warped image, $I\sp{\prime}$, and variance, $V\sp{\prime}$,  are related by a warping kernel, $k$, thus
\begin{equation}
I\sp{\prime}(x, y) = \sum\limits_{u,v} k(u, v)I(x - u, y - v)  
\end{equation}
\begin{equation}
V\sp{\prime}(x, y) = \sum\limits_{u,v} k(u, v)^2V(x - u, y - v) ,
\end{equation}
where $x$ and $y$ are image pixel indices and $u$ and $v$ are kernel pixel indices. Here the kernel has been normalised so it sums to unity. This warping process converts some variance into covariance, such that $V\sp{\prime}(x, y)$ no longer represents all of the noise associated with a pixel. To measure a warped pixel's total noise one needs to use a covariance matrix which accounts for the correlations between the pixels in the image. Storing the full covariance matrix would require a prohibitive amount of space so a much smaller matrix, known as the ``covariance pseudo-matrix'' is stored per stack image. 

The covariance pseudo-matrix, $\tilde{C}(i,j)$, describes the covariance of a single pixel with each of the pixels in its neighbourhood, with relative pixel coordinate
$(i, j)$. For initially uncorrelated data this matrix is simply a function of the warping kernel,
\begin{equation}
\tilde{C}(i,j) = \sum_{u,v} k(u - i, v - j) k(u,v) N 
\end{equation}
where $N=(\sum_{u,v} k^2(u, v))^{-1}$, such that $\tilde{C}(0,0)=1$ and $N=\sum_{i,j} \tilde{C}(i,j)$. The latter property follows from the normalisation of kernel, $k$. When making
measurements which combine many pixels the effect of covariance on the overall variance of the measurement can be approximated by simply boosting individual variances by
the factor $N$ and otherwise ignoring covariance. This approximation is asymptotically exact for apertures much larger than the kernel size. The value of $N$ changes from place to place on the sky but has an approximately Gaussian distribution with a mean of 1.379 with an rms of 0.006 for SAS2 \rps-band.  In Paper I we show how the warping process has no effect on the depth of images, but we will revisit the covariance pseudo-matrix in Section \ref{sec:masks}. 
\subsection{PS1 magnitudes, flags and nomenclature}\label{sec:psMags}
In this work we use Kron magnitudes \citep{1980ApJS...43..305K} as measured by the IPP code {\sc psphot} \citep{magnier2006} with zeropoints accurate to 10 mmag from the calibration described in \citet{eddie2} and \citet{tonry12}. Kron magnitudes measure flux in an aperture with a radius called the ``Kron radius'', which is some multiple (2.5 for PS1) of the first moment radius of the flux \citep{1980ApJS...43..305K}. Kron magnitudes are designed to contain the majority of flux for a given source profile regardless of size, but a small, profile dependent correction term is required to account for flux outside the Kron radius. For defining clustering samples we base our selection on uncorrected Kron magnitudes, as they are well defined for all of our objects. This correction needs to be considered when comparing total magnitudes from synthetic objects to observed quantities and when comparing to literature galaxy number counts. From Table 2 of Paper I we see this correction has a weak magnitude dependence, and changes by a few hundredths of a magnitude. It is also expected this correction will slightly depend on galaxy profile. For this work we adopt an average correction of $mag_{\rm{Total}} = mag_{\mathrm{Kron}} - 0.2$ to convert from Kron magnitude to total magnitudes; we will state explicitly wherever we apply this correction throughout this paper.  

For the purposes of star/galaxy separation we also use point spread function (PSF) magnitudes, which are magnitudes based on extrapolating the magnitude from a small aperture, chosen to maximise the signal-to-noise ratio (SNR), using the IPP PSF model (see Section \ref{sec:profiles}). We shall label these magnitudes with the suffix ``PSF'' to contrast with the Kron magnitudes which are labelled using the name of the filter, i.e. \grizy. 

All number count, colour-colour, colour-magnitude and clustering plots are corrected for galactic extinction using the dust maps and associated {\sc IDL} code of \citet{schlegel}, using the coefficients from \cite{sch2011}.  Star and galaxy separation and detection efficiency plots are all uncorrected for extinction, as the measured magnitude is more relevant for these plots. These extincted, measured magnitudes will be labelled with the suffix ``raw''.  

To remove known spurious detections we use IPP flags. All objects with the {\sc psphot} flags {\sc fitfail, satstar, badpsf, defect, saturated, cr\_limit, moments\_failure, sky\_failure, skyvar\_failure or size\_skipped} set are removed. Further discussion of these flags can be found in Paper I.                       

\subsection{SDSS magnitudes and flags for the comparison sample}\label{sec:sdss}
The SAS2 field overlaps with SDSS DR8 and is partially covered by the SDSS Stripe 82 co-added data \citep{annis11}, the size of the Stripe 82 overlap region is around 16 square degrees (see figure 1 of Paper I). We compare PS1 to both of these. Stripe 82 comparisons are particularly useful as Stripe 82 is deeper than PS1. 

SDSS measures magnitudes in an $\rm{asinh}$ magnitude system \citep{lupton}. We adjust this to the standard Pogson system using the formula available on the SDSS website\footnote[1]{http://www.sdss3.org/dr8; accessed 27/07/2012}. This adjustment is very small, at its maximum value, at $r=23.0$, it is only 0.04 magnitudes in size. The SDSS bands are slightly different to those of PS1, transformations are given in \citet{tonry12}. These transformations in our comparison band, \rps, are very small, less than 0.01 magnitudes for a wide range of colours in figure 6 of \cite{tonry12}, and hence are neglected.

\begin{figure}
 \includegraphics[width=3.3in]{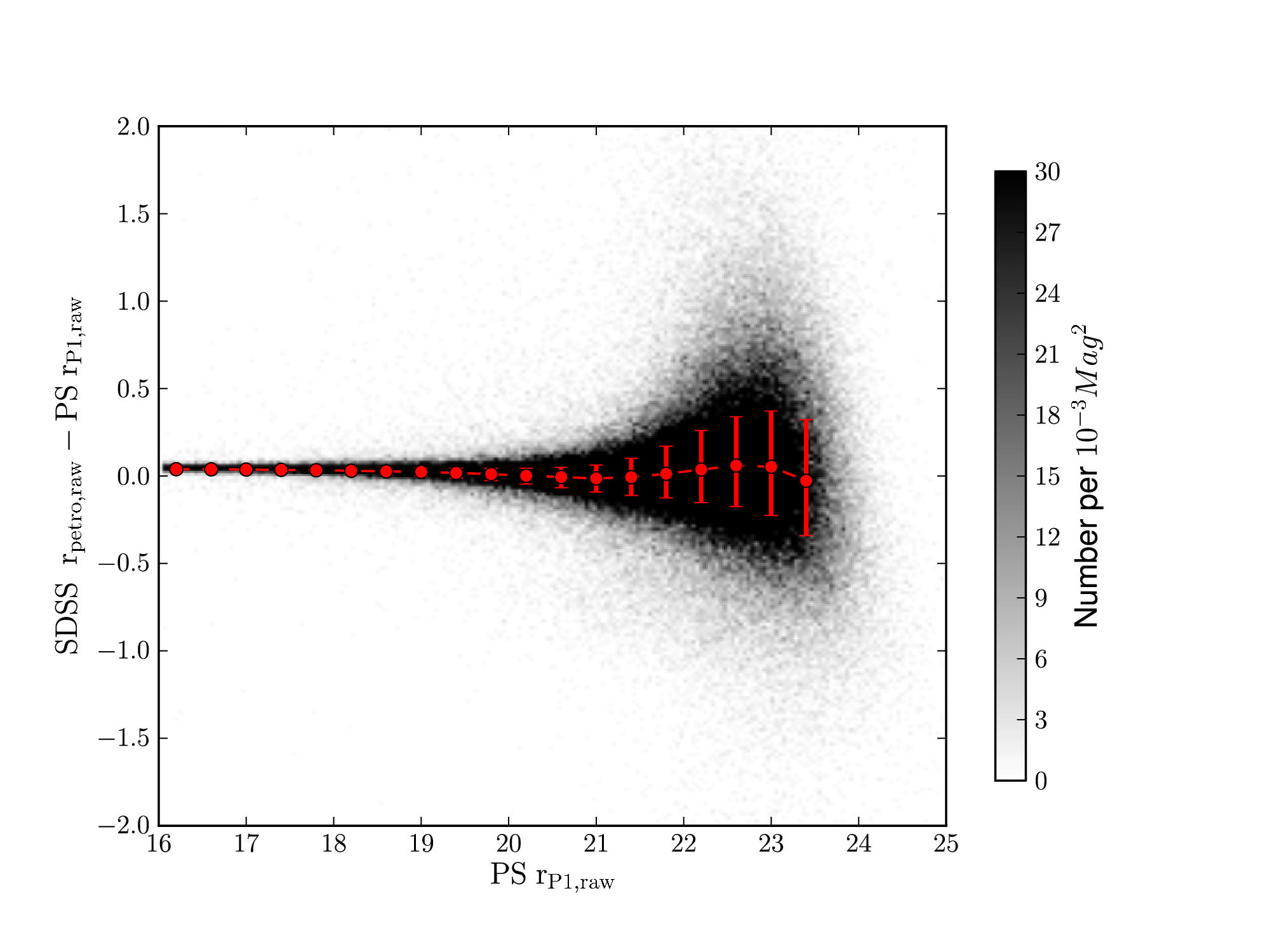}
 \caption{The difference between $r$-band SDSS Stripe 82 Petrosian magnitudes and \rps-band PS1 Kron magnitudes, for all objects in an overlap region. Points with error bars show the median values along with upper and lower quartiles. The two magnitudes are fairly well matched, with a small median offset that varies slightly with magnitude.}\label{fig:magmag}
\end{figure}

SDSS DR8 and SDSS Stripe 82 do not provide Kron magnitudes. Whilst the SDSS magnitudes measured using model fits, so called ``modelMags'',  give an estimate of the total magnitude of a galaxy, we want to select a magnitude estimator most similar to our Kron magnitudes (see Paper I for PS1 Kron and SDSS modelMag comparisons). Petrosian magnitudes \citep{petrosian1976}, a modified form of which are provided by SDSS \citep[see][]{blanton2001, yasuda2001} measure flux within an aperture of a size determined by the ratio of a surface brightness in an annulus around a source to the average surface brightness of the region interior to that annulus. In theory the fraction of flux enclosed by a Kron magnitude and a Petrosian magnitude could differ. A comparison of PS1 measured Kron magnitudes and SDSS DR8 Petrosian magnitudes (Fig. \ref{fig:magmag}) shows that these two magnitude measures are fairly well matched in the \rps-band and $r$ band for objects in SDSS. 

To define SDSS galaxies we use the \cite{strauss02} star-galaxy separator, 
\begin{equation} \label{eq:starGal}
r_{\rm{psf}} - r_{\rm{model}} > 0.3 ,
\end{equation}
where $r_{\rm{psf}}$ is the SDSS PSF magnitude and $r_{\rm{model}}$ is the SDSS model magnitude. We use SDSS flags to remove false positives in SDSS DR8. Following the spectroscopic target selection of \cite{strauss02} we reject SDSS objects with {\sc saturated} or {\sc bright} flags, and require the {\sc binned1} flag to be set (i.e. a $5\sigma$ detection). Again following \cite{strauss02} we apply, to the DR8 data, a Petrosian half light surface brightness cut of
\begin{equation} \label{eq:surfaceB}
\mu_{50} = m_{\mathrm{petro}} + 2.5\mathrm{log_{10}}(\pi R_{\mathrm{ 50, petro}}^2) <  24.0 ,
\end{equation}
where $m_{\mathrm{petro}}$ is the Petrosian magnitude and $R_{50, \rm{petro}}$ is the radius enclosing 50\% of the Petrosian flux. \cite{strauss02} adopted a similar cut to remove low surface brightness false positives; though they used a slightly more complicated cut than ours, which was dependent on sky values and fibre magnitudes. We adopt this simplified, less conservative cut (the \cite{strauss02} could be as bright as $\mu_{50}<23.0$) to DR8 as we find it is sufficient to remove SDSS false (unmatched to PS1) detections from the magnitude ranges we consider. Applying this surface brightness cut limits SDSS DR8 depth faintward of $r=20.0$, so we do not compare to SDSS DR8 faintward of this value. With more work it is likely possible to measure SDSS DR8 clustering over SAS2 for galaxies fainter than this, but we choose instead to use literature and Medium Deep data for faint clustering comparisons. 

We do not apply any surface brightness cut to Stripe 82 data as our main use of Stripe 82 is to estimate PS1 depth and these cuts could limit Stripe 82 depth. How Stripe 82 false detections affect this work will be discussed in Section \ref{sec:detectionEff}. Stripe 82 does not have a publicly available mask for the co-added data, so we created our own by visual inspection of the area. This masks defines areas with no Stripe 82 imaging and removes a satellite trail in Stripe 82. 

\begin{figure}
 \includegraphics[width=3.6in]{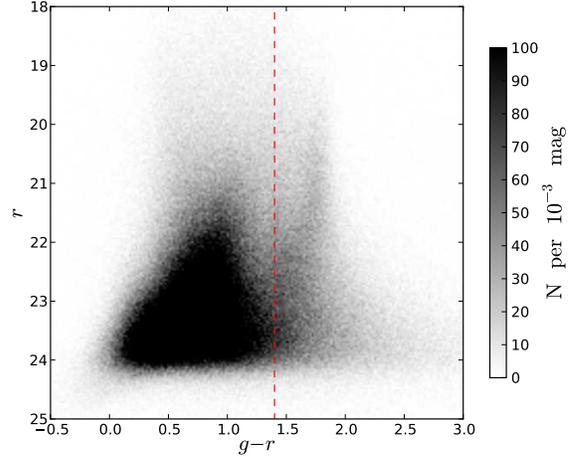}
 \caption{A  colour magnitude diagram of Stripe 82 galaxies, using Stripe 82 apparent model magnitudes. The red dashed line marks our separator between red and blue galaxies.}\label{fig:stripe82CMD}
\end{figure}

A further use for Stripe 82 is to test how strongly detection efficiency depends on apparent colour. A galaxy's colour is correlated with its morphology, red galaxies tend to be ellipticals and blue galaxies tend to be spirals. Galaxies with different morphologies have different surface brightness distributions and as such may have a different chance of being detected. Since galaxy clustering is a function of colour and morphology, with red ellipticals being more clustered, this effect could modify our clustering for cuts and depth corrections based on apparent magnitude. Fig. \ref{fig:stripe82CMD} shows a colour magnitude diagram using Stripe 82 model magnitudes for objects classed as galaxies by Stripe 82's own morphological star and galaxy separator, {\sc type = 3}. We separate galaxies on the red sequence from those in the blue cloud using the cut indicated on  Fig. \ref{fig:stripe82CMD}, $(g - r) = 1.4$. We will use this sample of red and blue galaxies when testing the dependence of detection efficiency on apparent colour and hence morphology.

\subsection{PS1 Medium Deep Data}\label{sec:psMD}
When comparing our faint galaxy clustering to other measurements we both compare to literature data and to results from the much deeper and more spatially homogeneous PS1 Medium Deep survey. Foucaud et al. (in preparation) has produced their own stacks of Medium Deep field 7 (MD07) using PS1 data and reduced them using {\sc SExtractor} \citep{bertin}. They measure the Kron magnitudes of galaxies, using {\sc SExtractor MAG\_AUTO}, and star/galaxy separate using a combined morphological and SED fitting approach. They also adopt a mask to remove bright stars and poorer quality data. After masking, MD07 has an area of 7 square degrees, much smaller than the SAS2 field. For more details on 
these stacks see \citet{jian2013} and Foucaud et al. (in preparation).

\section[Masks]{Angular Masks and False Positives}
\subsection{Creating the mask}\label{sec:masks}

To create a set of random points suitable for measuring clustering and to remove regions of low data quality we define a new set of angular masks. These masks differ from IPP image masks in that a single, unique mask covers the whole region of interest. In IPP two overlapping skycells will have two different masks, one for each skycell. As well as masks we produce variance maps and coverage maps binned-up to the same resolution as our mask pixels. We take variance maps, coverage maps and image masks at the native pixel scale and compute their mean on a grid of $12000^2$, $3.3\arcsec \times 3.3\arcsec$ equal area pixels, which covers the whole SAS2 area. As binned-up pixel boundaries do not align with the IPP pixel boundaries, we assign pixels to their nearest binned-up pixels. Our binned up pixel grid has the same rotation as the IPP pixels. For our coverage maps we take the lowest value of any IPP pixel in our binned-up pixels, to be conservative in our estimates of low coverage areas. Our binned-up pixel size was chosen to preserve the fine structure in the variance whilst still yielding a mask of manageable size. Experimenting with different mask and map pixels sizes and different mask and map tessellations is left for later work. 

Only taking into account the variance recorded in the variance maps would result in underestimating the noise, as we would be ignoring the covariance. We therefore multiply variance values from IPP variance maps by the sum of the elements of their associated covariance pseudo-matrix (see Section \ref{sec:data}). This is almost the same as multiplying all of variance map values by a constant, as the rms of this scaling factor, given in Section \ref{sec:data}, is only around $0.5\%$ across the SAS2 field. We carry this scaling out to allow easier comparison to the work of Paper I, which works with uncorrelated noise measurements. We also apply this scaling now as it could become more important if the warping kernel were to change.   

Where data from two skycells overlap we take data from the skycell whose centre is closest to the overlapping data. We do this for both the pixels and the object detections to ensure the catalogues, masks and maps are consistent.

As well as defining the basic geometry of the survey, we also use angular masks to avoid two other types of potential problem: deblending and image artifacts. 

\subsection{Masks for bright stars}
 In common with a large amount of image reduction software \citep[see e.g][]{bertin}, {\sc psphot} can split bright objects and diffraction spikes into multiple detections. To avoid this we mask out regions around bright stars. To decide mask sizes we use photometry from the UCAC4 catalogue \citep{z13} rather than PS1, since PS1 saturates at around $\rps < 15.0$. We use $R$-band photometry from the UCAC astrograph up to a bright limit of  $R=10.0$, where the astrograph becomes saturated. To mask even brighter objects we use $V$-band data from Hipparcos, FK6 and Tycho-2. This data is already included in the UCAC4 catalogue. \cite{z13} states that the UCAC4 catalogue is a complete catalogue of stars down to $R<16$. 

\begin{figure}
\begin{center}
 \includegraphics[width=3.1in]{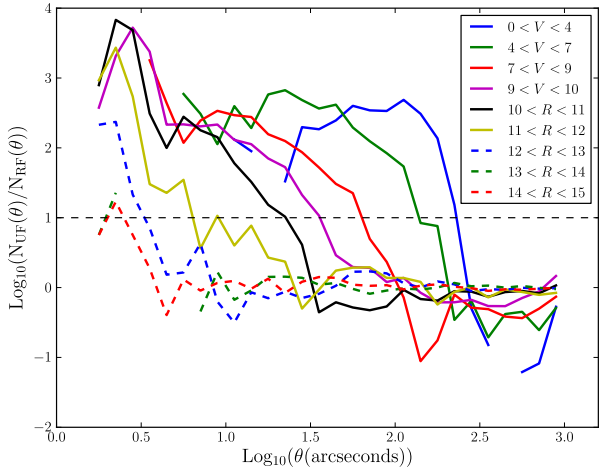}
 \includegraphics[width=3.68in]{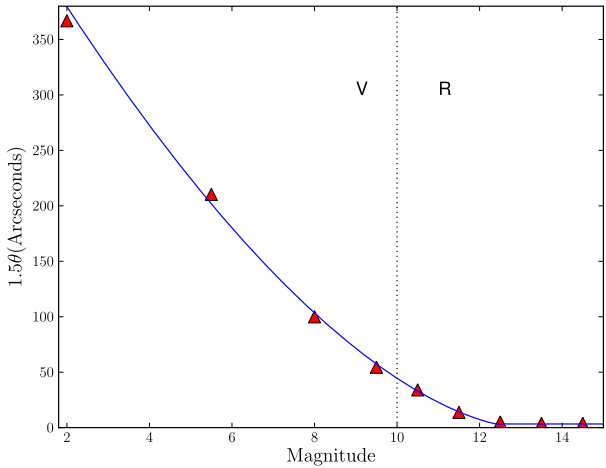}
 \caption{\emph{(Top)} The correlation of false positive detections with bright stars in the UCAC4 catalogue. The lines show the ratio of the number of false to UCAC4 pairs to the number of false to random pairs as a function of R and V magnitudes from the UCAC4 catalogue. Using two different bands is necessary as the astrograph measuring R magnitudes saturates for very bright stars. We see a clear correlation between false positives and UCAC4 sources. The level at which there are 10 times as many UCAC4 to false pairs as random to false pairs is marked with a horizontal dashed line. \emph{(Bottom)} The largest separation corresponding to this level for each bin, multiplied by 1.5. A fit to these points (blue curve) sets the size of the bright source mask as a function of R and V magnitude.}\label{fig:maskSize}
\end{center}
\end{figure}

We identify likely candidates for false positives by identifying objects in the $\rps$-band that are not in the $\ips$-band catalogue, with a $0.5\arcsec$ matching radius. To eliminate objects that are not detected in both bands due to image depth, we remove objects with $\rps > 20.0$. We assume these candidate false positives trace the spatial distribution of all false positives caused by bright stars. Selecting the central, deeper region we count ``false positive'' and UCAC4 pairs as a function of angular separation, $FU(\theta)$, as well as pairs of ``false-positive'' and random points uniformly distributed across the area, $FR(\theta)$. We calculate the ratio of these pairs
\begin{equation}
 \frac{N_{\rm{FU}}(\theta)}{N_{\rm{FR}}(\theta)} = \frac{FU(\theta)}{FR(\theta)}\frac{n_{\rm{R}}}{n_{\rm{D}}}
\end{equation}
where $n_{\rm{D}}$ is the number of UCAC4 objects and $n_{\rm{R}}$ is the number of random points. This technique is very similar to a cross-correlation function. We adopt this technique to map out the scale out to which one finds false positives around bright stars. In Fig.~\ref{fig:maskSize}(top) we plot the results as a function of UCAC4 $R$ magnitude and $V$ magnitude. The brightest bin contains only one $V=2.33$ magnitude star. From Fig.~\ref{fig:maskSize}(top) we can see brighter objects cause false positives out to a larger spatial extent than fainter ones. We also see a relative deficit of false positives at smaller separations. This is due to masked, saturated regions closer into the bright object. Also note that false positives are preferentially found near brighter objects all the way down to the magnitude limit of $R=15.0$. Whilst Fig.~\ref{fig:maskSize}(top) shows one is ten times more likely to find false positives at a separation of $~3\arcsec$ from objects with $14<R<15$, it does not imply that all of these objects cause false positives and in real terms the number of bright false positives is very small. To decide on the size of mask to put on bright objects, as a function of $R$ and $V$ magnitude, we use the last crossing of the $\log10(N_{FU}(\theta)/N_{FR}(\theta))=1.0$ line as a reference separation and increase this distance by 50\%. The curve describing mask size is smooth across the $V$-band to $R$-band boundary, see Fig.~\ref{fig:maskSize}(bottom). We fit these sizes with a simple power law, truncated such that mask size cannot be less than one mask pixel (i.e. $3.3\arcsec$),

\begin{equation}
     r_{\mathrm{mask}} =\left\{ 
     \begin{array}{ll}
     7.26(13.0 - m)^{1.65}   & \mbox{if $r_{\mathrm{mask}} \geq 3.3$} \\
     3.3                     & \mbox{otherwise}  \\
    \end{array}\right.
\end{equation}
where $r_{\rm{mask}}$ is the mask radius in arcseconds, and $m$ is the stellar magnitude. We use this to mask down to $R<15$ and $V<10$. 

\subsection{Masks for regions of low quality data}
\begin{figure}
\begin{center}
 \includegraphics[width=3.5in]{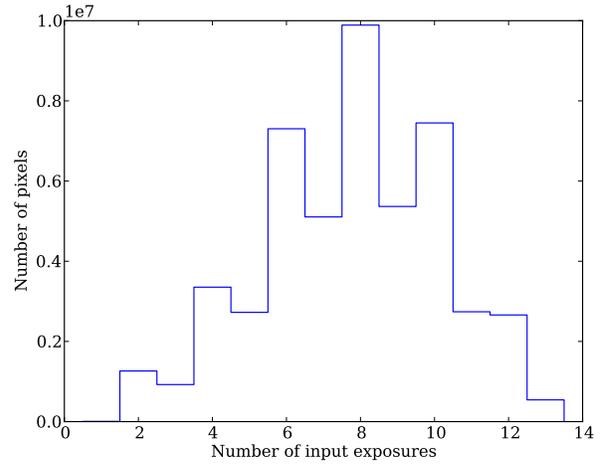}
 \caption{A histogram of the lowest coverage values, i.e. fewest exposures per stack pixel, in each of our binned up coverage map pixels.}\label{fig:covHist}
\end{center}
\end{figure}

The second potential issue we combat with masks is that certain regions of PS1 images have instrumental signatures (i.e. image artifacts) caused by scattered light and electronic noise. This is particularly noticeable in regions of low coverage where we do not have sufficient numbers of exposures to remove these image defects statistically, i.e. by median filtering or outlier rejection in the stacking procedure. We therefore mask regions with a coverage of three exposures or fewer. In the finished survey the area with coverage this low should be very small. To estimate this value we took the central area, $331.0<\alpha\rm{(deg.)}<336.0$ and $-3.0<\delta\rm{(deg.)}<3.0$, of our binned up version of the coverage map and produced Fig.~\ref{fig:covHist}. The central area of SAS2 should be representative of the finished $3\pi$  data and as such we can see from Fig.~\ref{fig:covHist} only $4\%$ of the full survey area should be lost by this cut.
Masked regions are expanded by a one binned-up mask pixel border in order to exclude from the catalogue objects with unreliable measurements caused by being on the edge of the mask. This is similar to using cuts in the IPP value {\sc psf\_qf\_perfect}, which quantifies the fraction of masked or suspicious pixels in a source (for more details on these cuts see Paper I).

\subsection{The effects of masking}

\begin{figure*}
\begin{center}
 \includegraphics[width=6in]{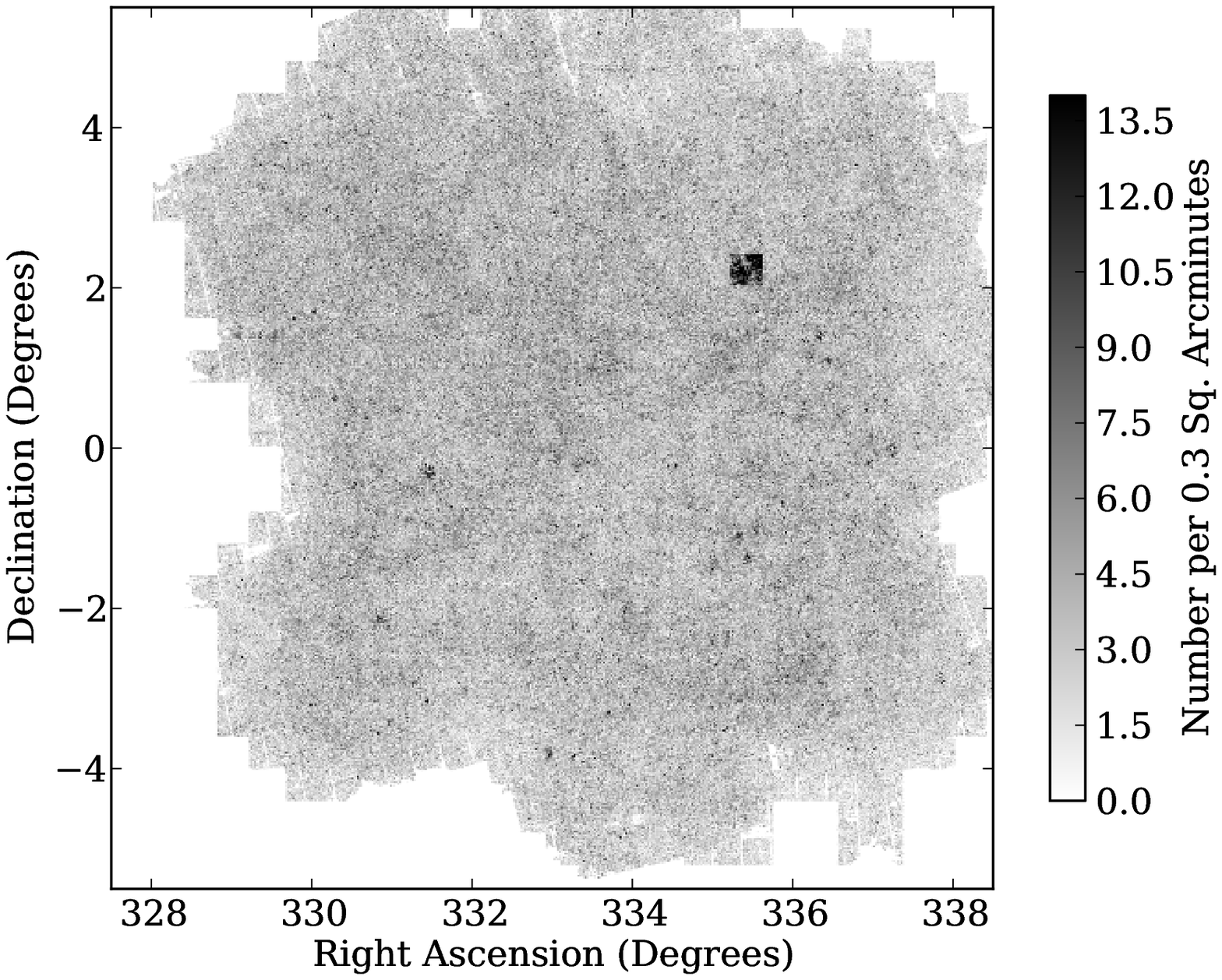}
 \includegraphics[width=6in]{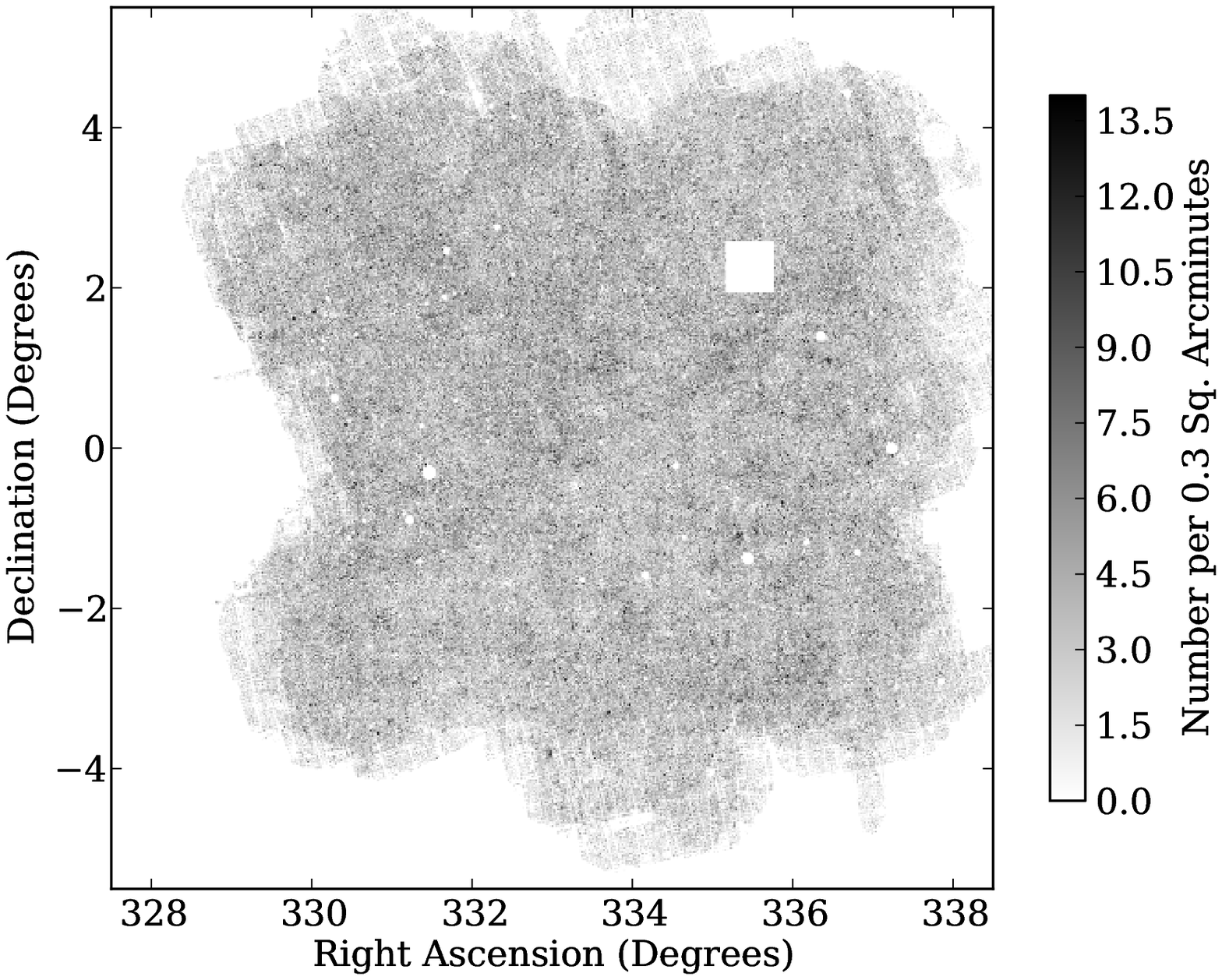}
 \caption{\emph{(Top)} A plot of all detections in SAS2, binned into 0.3 square arcminute pixels. \emph{(Bottom)} The same plot after masking and applying the flags specified in Section 2.2. We can see the circular star masks, the areas near the edge masked due to our cut on low coverage and the square area masked by hand where the data reduction failed. Over-densities caused by stars are removed, the remaining darker regions are caused by variable image depth or genuine over-densities in the object distribution.}\label{fig:mask}
\end{center}
\end{figure*}

The source detections before and after applying the final mask are shown in Fig.~\ref{fig:mask}; the regions of fewer objects on the outskirts of the masked field are not caused by depth variations but simply the larger number of masked pixels caused by a lower coverage in these areas. The grid like patterns are also caused by our masking of low coverage regions; the grid pattern in coverage is caused by gaps between individual chips on the detector. One can see from Fig.~\ref{fig:mask} how our angular mask removes peaks of false positives caused by bright objects: peaks of false detections in the unmasked field are removed in the masked field. Finally we mask, by hand, a square region in SAS2 where the data reduction process failed, an issue that will be rectified for the final survey.

\begin{figure}
\begin{center}
 \includegraphics[width=3.5in]{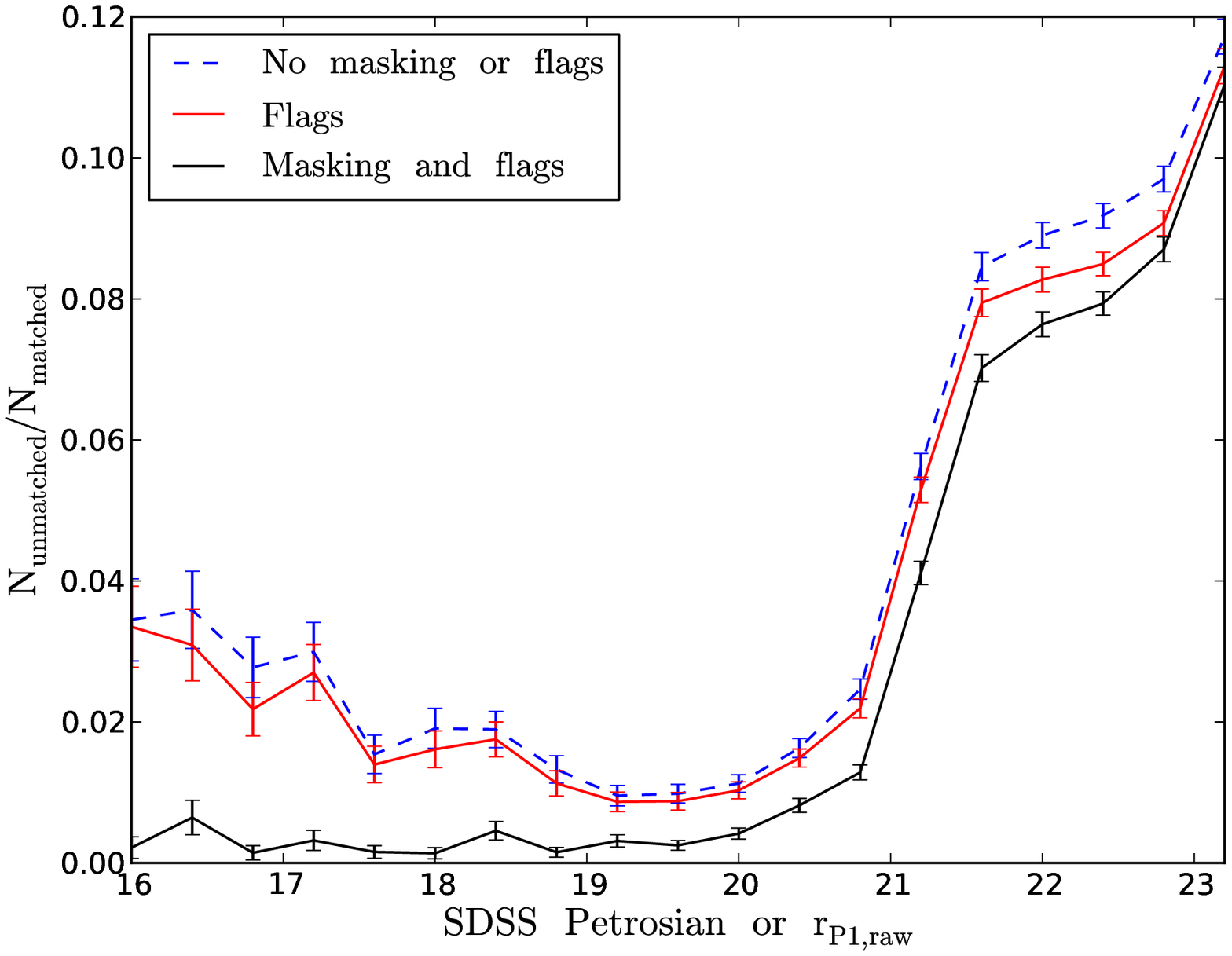}
 \caption{The fraction of unmatched objects as a function of magnitude, error bars show Poisson noise. The improvement gained from applying the flags and applying the masking is clear. }\label{fig:hist}
\end{center}
\end{figure}

A quantitative measure of how our mask removes false detections was made by cross matching the Stripe 82 and PS1 catalogues after applying our SDSS Stripe 82 mask (see Section \ref{sec:sdss}) to PS1 data and the PS1 mask to Stripe 82 data. Fig.~\ref{fig:hist} shows the fraction of unmatched objects to Stripe 82, for an $\sim 8$ square degrees overlap region and a matching radius of $1\arcsec$, before and after applying the masking and flags. 

Fig.~\ref{fig:hist} shows a decrease in the fraction of false positives once flags have been applied and masking conducted. In particular brighter false positives associated with bright stars are almost entirely removed. Some unmatched objects do remain, but at magnitudes brighter than $\sim21$ these are mostly real objects missed by Stripe 82 or objects with proper motions. Fainter than this false positives can be caused by the previously mentioned instrumental signatures. Note that Paper I achieves similarly low numbers of false positives by applying the {\sc psf\_qf\_perfect} flag; however the use of this flag, which depends on the number of masked or suspect pixels near a source, can change the angular selection function. The approach using masks presented here deals with these false positives in a way that keeps track of this, which is more appropriate for clustering studies. 

\section[Star/Galaxy Separation]{Star/Galaxy Separation}
\subsection{Galaxy and stellar profiles}\label{sec:profiles}
To describe our morphological approach to star galaxy separation, we first review the basic properties of galaxies. Galaxies have light profiles well fitted by the famous S{\'e}rsic functions \citep{sersic}. For a review see \cite{Graham05}. In flux this can be expressed as
\begin{equation}\label{eq:sersic}
F(R) = F_{\mathrm{eff}}  \exp  \left(  -b_{n} \left[ \frac{R}{R_{\mathrm{eff}}} \right]^{\frac{1}{n}} - 1 \right) ,
\end{equation}
where $R$ is the distance to the centre, $ F_{\rm{eff}}$ is the flux at $R_{\rm{eff}}$ and $b_{n}$ is a scaling constant that depends on the index, $n$, defined such that  $R_{\rm{eff}}$ is the half light radius. A value of $n=1$ and $b_{n} = 1.678$ gives an exponential profile, typical of the discs of spiral galaxies while a value of $n=4$ and $b_{n}=7.669$ gives the de Vaucouleurs profile typical of elliptical galaxies \citep[see][]{Graham05, deVaucouleurs}. Due to the atmosphere and telescope optics galaxy light profiles appear convolved by the point spread function (PSF) of the instrument and the atmosphere. In the PS1 IPP stars are fitted with a PSF model of the form 
\begin{eqnarray} \label{eq:ps1v1}
I=\frac{I_{0}}{1 + kz + z^{3.33/2}}, \\
\mbox{where } z = \frac{x^2}{2\sigma_{x}^2} + \frac{y^2}{2\sigma_{y}^2} + xy\sigma_{xy} ,
\end{eqnarray}
where $I_{0}$ is the central intensity, $x$ and $y$ are the $x$-axis and $y$-axis distances from the centre, $k$ is a free parameter and $\sigma_{x}$, $\sigma_{y}$, $\sigma_{xy}$ are free parameters that represent the $x$-axis width, the $y$-axis width and a cross term respectively. During image reduction IPP fits the PSF model parameters on a grid across each skycell. Between these grid points parameters are interpolated to give a smoothly varying model of the PSF. Typically PS1 PSFs, and indeed real PSFs in general, have more extended wings than Gaussian PSFs of the same full width half max (FWHM): figure 5 of Paper I gives a typical curve of growth for a PS1 PSF.   

\subsection{Synthetic objects}\label{sec:fakes}
To develop and test morphological star/galaxy separation we generate synthetic sources with the profiles as described in Section \ref{sec:profiles}. To generate a synthetic star one needs to simply choose a magnitude and a position and then evaluate the model. Generating a galaxy is harder as several parameters must be chosen, namely: the position, the bulge to disc ratio, the S{\'e}rsic index, the size, the ellipticity and the orientation on the sky. The last of these is chosen at random. As the clustering of the synthetic sources is not important here a position is randomly assigned. 
 
When choosing the S{\'e}rsic index we approximate the Universe as being made up entirely of de Vaucouleurs profiles for elliptical type galaxies and bulges,
or exponential profiles for discs. This follows the classic bi-modality in S{\'e}rsic index between elliptical galaxies and discs. In reality galaxies follow a distribution of S{\'e}rsic indices, with elliptical galaxies displaying a positive correlation between luminosity and S{\'e}rsic index \citep[see e.g.][]{Ferrarese2006}. For this work the galaxies that will be difficult to star/galaxy separate will have small angular sizes, faint apparent magnitudes and are convolved with a PSF so we feel this approximation makes negligible difference to our results. We also treat bulges in disc galaxies in the same way as elliptical galaxies, which is a common approximation adopted in the literature \citep{bertin, shen}. 

For the axis ratios of discs we choose a random inclination angle, $i$, distributed uniformly in $\cos(i)$ and assuming circular flat discs with a thickness which is some fraction, $t$, of the radius we calculate the apparent axis ratio, $e_{\rm{sky}}$, using simple geometry as
\begin{equation}
  e_{\mathrm{sky}} =  \cos(i) + t\sin(i)   .
\end{equation}
We take $t=0.1$ for our disc height to radial scale length ratio. The resulting distribution is flat and a reasonable fit to the observations in \citet{padilla08}. For bulges we select a major to minor axis ratio, $e$, between 0.3 and 1.0, corresponding to the classical elliptical types of E0 to E7 \citep[see e.g.][]{mo}. Within this range we select $e$ from a truncated Gaussian distribution of mean $\mu=0.75$ and variance of $\sigma^2=0.1$, which we chose to give a reasonable fit to the data in figure 4 of \citet{padilla08}.

For physical galaxy sizes we use the empirically measured relation and its scatter given in equations 14, 15 and 16 of \citet{shen}. We adopt parameters measured in \citet{shen} for galaxies separated into late and early types by S{\'e}rsic index (figure 6 of that paper). It was reported in \citet{dutton2011} that using the \citet{shen} measurements would result in discs too small by a factor of around 1.4, due to not factoring in the effects of inclination which decreases the size by the square root of the apparent axis ratio. We therefore increase the size of our disc galaxies by this factor. We also correct the empirical bulge size relation for this effect, adopting a correction of 1.2, calculated from the typical bulge ellipticity $\mu$ . For bulges and elliptical galaxies we choose not to extrapolate the relation from \citet{shen} to fainter magnitude bins than measured in that paper. Instead, we keep the sizes of bulges and elliptical galaxies fixed fainter than $M_{r}=-19$; this is motivated by observations that dwarf elliptical galaxies have a nearly constant size regardless of magnitude \citep[see e.g.][]{shen,mo}. 

We now have a relation between physical size and absolute magnitude, therefore we need a redshift and an absolute magnitude to predict angular sizes. One could generate these using observed luminosity functions and redshift distributions, but here we use data from the mock catalogues produced for \citet{merson} using the galaxy formation model presented in \citet{bower06}. Using these catalogues gives us the potential to extend this work to generate synthetic images with realistic galaxy clustering. For the purposes of this work, however, we use random angular positions. The model adopts a concordance cosmology of $\Omega_{\rm{m}}=0.25, \Omega_{\Lambda}=0.75, \Omega_{\rm{b}}=0.045, h=0.73$; we use this cosmology for the whole of this work. The galaxy formation model gives magnitudes and redshift distributions in good agreement with observations at low redshift \citep{bower06}. We split the total flux of the model galaxy into a bulge component and a disc component by randomly sampling bulge to total ratios from table 3 of \cite{simard}, which gives an observational estimate of bulge to total ratios for around a million SDSS galaxies. The measured magnitudes of the synthetic galaxies are faded by the mean extinction of SAS2, 0.2 magnitudes (Paper I).

Once we have the galaxy morphological properties, we evaluate Equation \ref{eq:sersic} on a pixel grid of a linear scale three times smaller than PS1 warped pixel scale of $0.25''$ before binning up. This is to minimise the effect of gradients in the profile across pixels. Pixels on the finer grid whose centres are closer than $0.1''$ to the profile centre are further subdivided 3 by 3 to take into account the steeper profile near the centre. If any of these subdivided fine pixels are on the centre of a de Vaucouleurs profile, an analytic integral is used to approximate the flux required, as de Vaucouleurs profiles asymptote to infinity at zero. Stars, conversely, are evaluated directly on to the native pixel scale as this is the scale at which the model is measured. Galaxy profiles are convolved with the PSF using the C-library {\sc fftw} \citep{FFTW05}. The grid dimensions are chosen to ensure that the finished, convolved galaxy image contains more than 99.8\% of the flux. Stars are evaluated on a grid of $36\arcsec$ by $36\arcsec$ which contains more than 99.9\% of the flux for PS1 SAS2 PSFs. 

Paper I shows results from our synthetic stars agree with a set of synthetic stars produced by IPP. It also uses our synthetic objects to test PS1 depth and photometry. Interested readers can refer to Paper I for basic results from the synthetic objects, such as recovered versus input magnitude. 

\subsection{Morphological Separator}
\label{sec:sgSep}

\begin{figure*}
\begin{center}
  \includegraphics[width=6.5in]{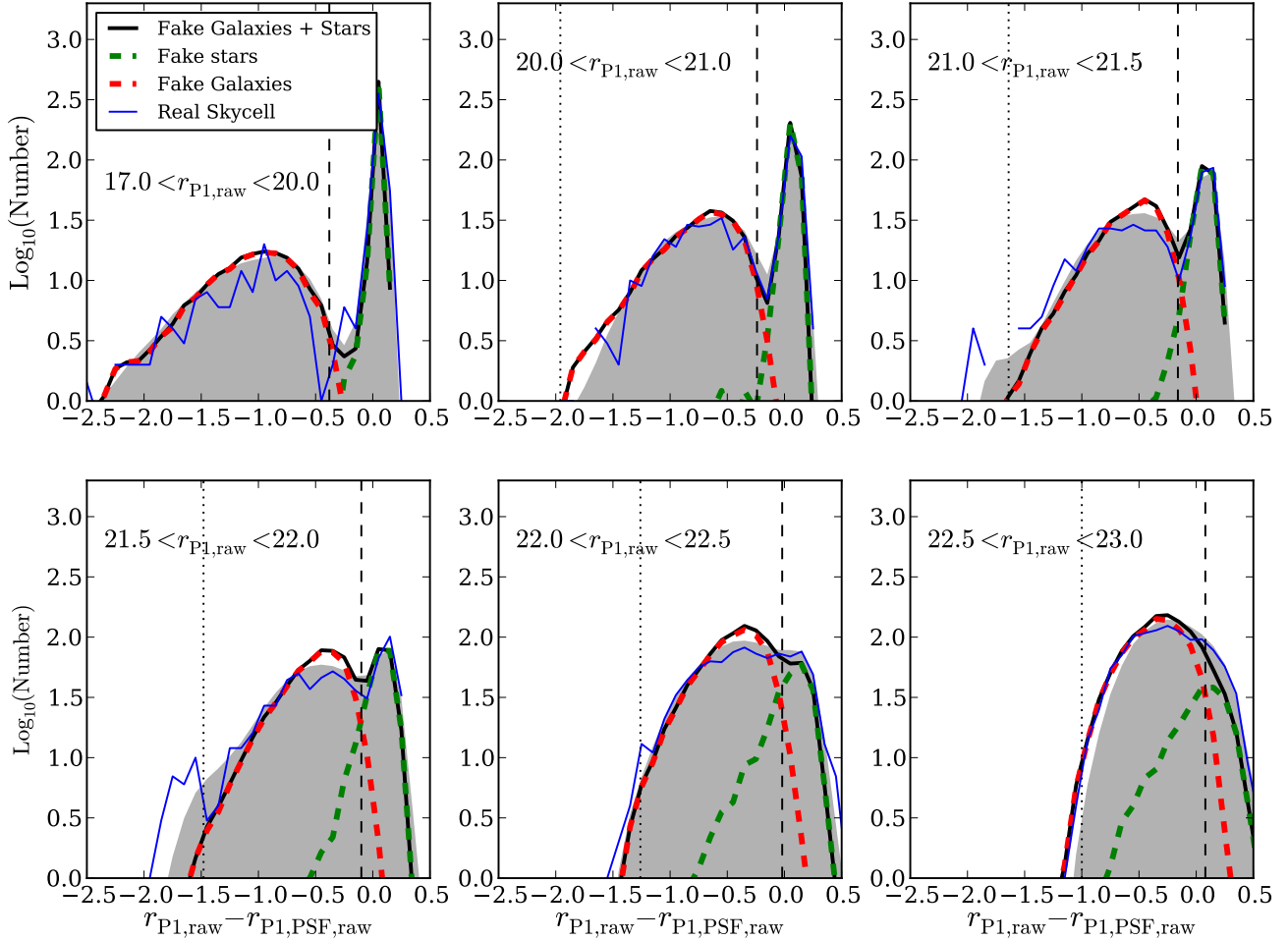}
  \caption{Kron minus PSF \rps-band magnitudes for all synthetic objects (black), synthetic stars (green dashed) and synthetic galaxies (red dashed) placed into the SAS2 skycell 1315.028, which has a PSF FWHM typical of SAS2 data. Also plotted are the real sources from that skycell (blue) and all sources in SAS2 (grey shaded area), the latter is normalised to the area of skycell.1315.028. The vertical dashed line shows the position of the star and galaxy separation cut, the dotted vertical line shows the position of the extreme Kron minus PSF magnitude cut.}\label{fig:sgHist}
\end{center}	
\end{figure*}

The PS1 SAS2 \rps-band skycell 1315.028 was taken as an example and 286 synthetic galaxies and 300 synthetic stars down to a limit of $\rps \le 23.5$ were inserted into it, created as described by Section \ref{sec:fakes}. This skycell was chosen as it has a PSF FWHM typical of SAS2. The PS1 photometry code {\sc psphot} was run on this skycell and this process was repeated 40 times yielding data from 11,440 synthetic galaxies and 12,000 synthetic stars. Motivated by the often used star and galaxy separator of a PSF magnitude minus an aperture-like magnitude \cite[e.g.][]{strauss02}, we show in Fig.~\ref{fig:sgHist} a histogram of the {\sc psphot} measured Kron minus PSF magnitude for the synthetic galaxies, the real sources in this skycell and for sources over the whole of SAS2. The number of synthetic galaxies and stars are scaled to the observed number of objects in each magnitude bin. We can see from Fig.~\ref{fig:sgHist} that the synthetic stars and galaxies follow the distribution of the real sources. This indicates we are justified in using Kron minus PSF magnitude as a star and galaxy separator. We see the synthetic stars follow a peaked, stellar locus whereas the synthetic galaxies follow a more negative locus of extended sources.  

\begin{figure}
\begin{center}
 \includegraphics[width=3.3in]{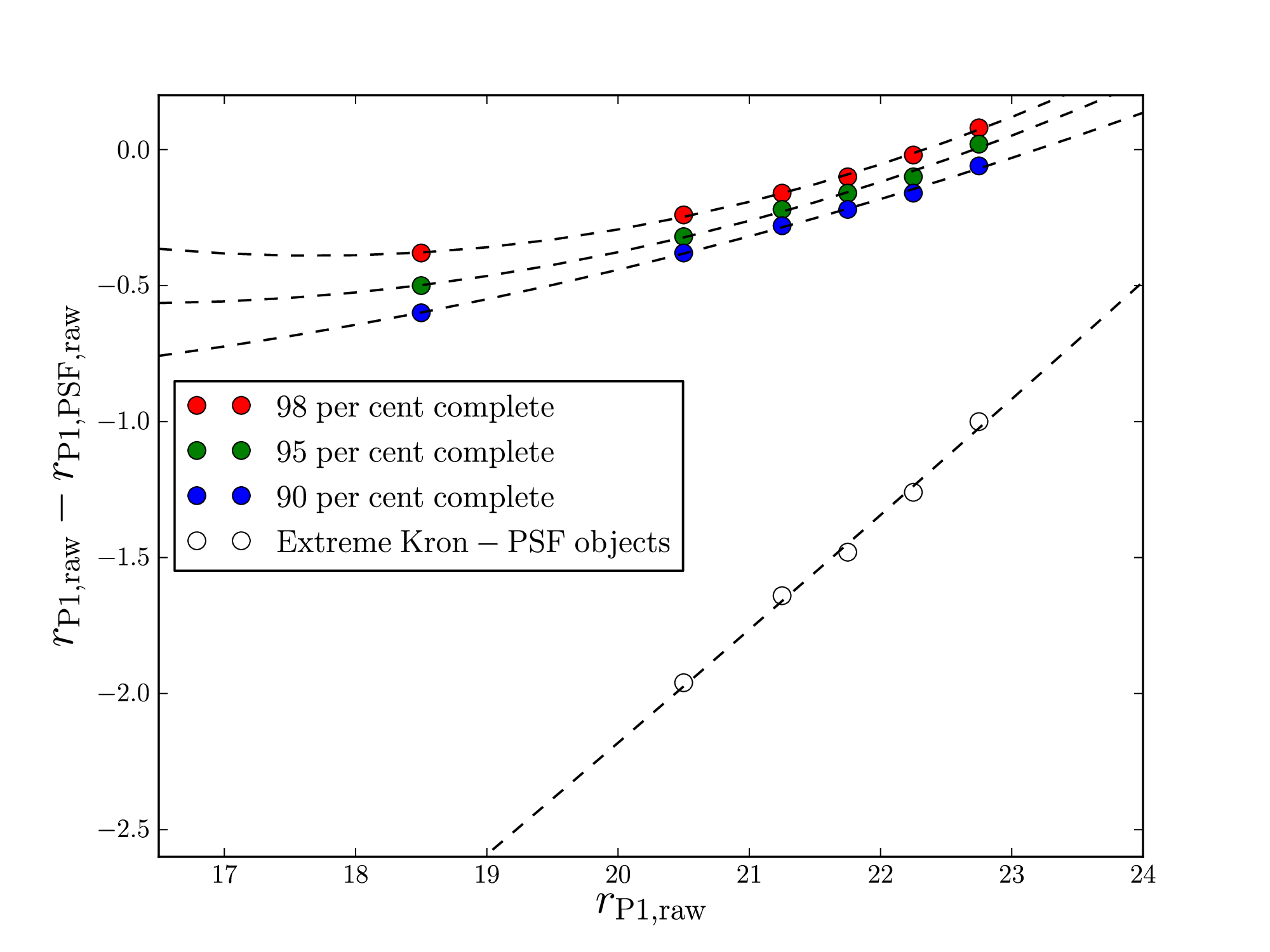}
 \caption{Galaxy (filled points) and extreme Kron minus PSF cuts (black circles) in the $r_{\rm{P1,raw}} - r_{\rm{P1,PSF,raw}}$ versus $r_{\rm{P1,raw}}$ plane, with colours indicating their completeness as found by the simulations shown in Fig.~6. The points are fitted with a second order polynomial (Eq. \ref{eq:sgCut}).}\label{fig:sgCuts}
\end{center}
\end{figure}

\begin{figure}
\begin{center}
 \includegraphics[width=3.6in]{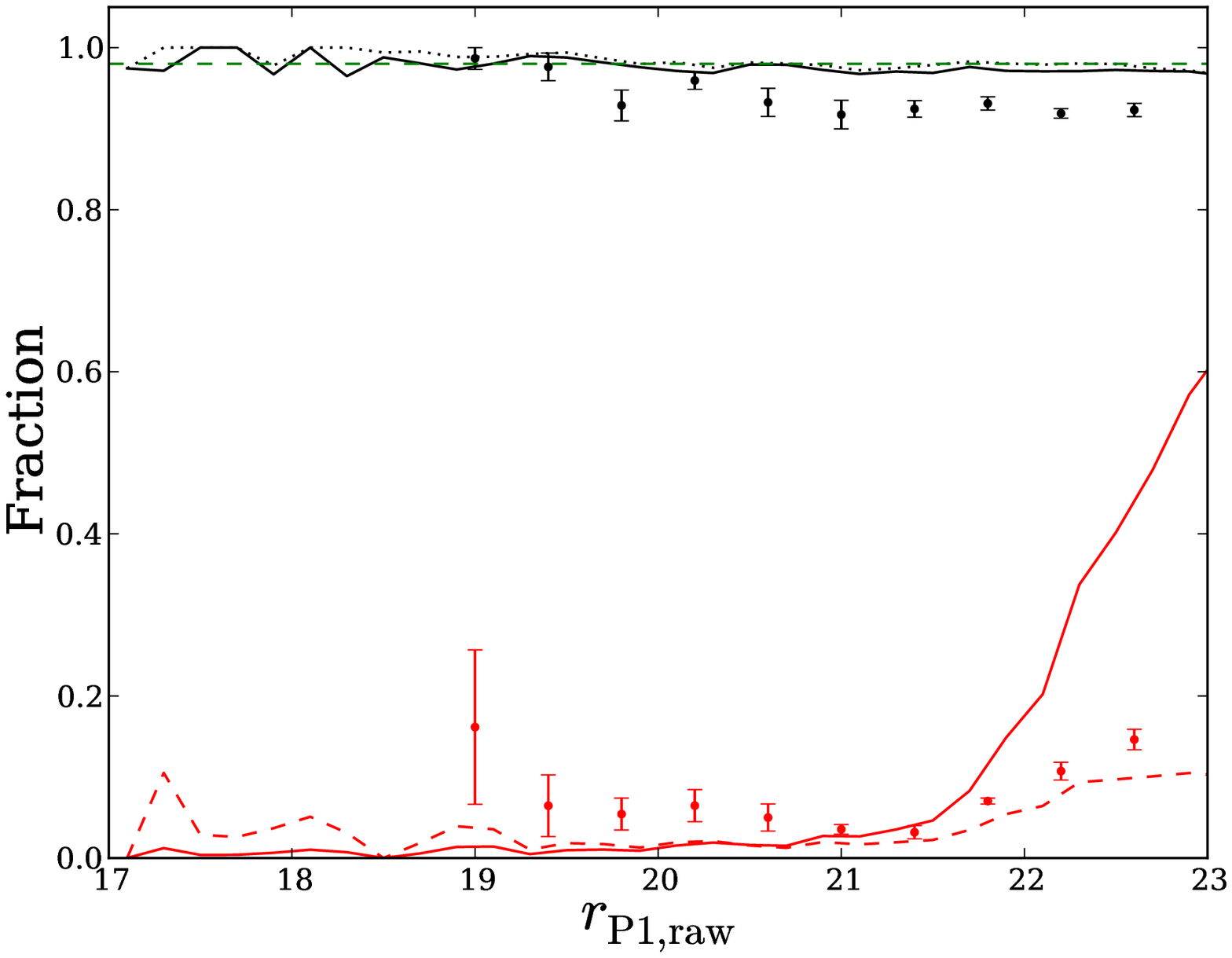}
 \caption{The probability of correctly classifying a source as a galaxy using the 98\% cut (black) and the probability of misclassifying a star as a galaxy (red solid), as predicted using our synthetic objects in Fig.~\ref{fig:sgHist}. The dotted line is for galaxies before the application of the extreme Kron minus PSF cut. The dashed green line shows the target completeness of 98\%. Also plotted is the predicted amount of stellar contamination as a fraction of the 98\% galaxy sample (red dashed), found from scaling the probability of misclassifying a star with power law fits to the bright end of the observed star and galaxy number counts. The points with error bars are estimates based on our comparison to the spectroscopic classifications of VVDS sources, as explained in Section 3.3.2.}\label{fig:sgCompl}
\end{center}
\end{figure}

We use our synthetic objects to define cuts in Kron minus PSF magnitude ($\Delta_{\rm{kron-psf}}$ hereafter) that define samples of stars or galaxies. We can also define a smallest allowed value of $\Delta_{\rm{kron-psf}}$ for galaxy samples, this removes objects with extremely negative $\Delta_{\rm{kron-psf}}$ which are likely false positives. We place this extreme $\Delta_{\rm{kron-psf}}$ cut at a value where only $0.5\%$ of synthetic galaxies are to the left of this cut. Fig.~\ref{fig:sgCuts} shows cuts in $\Delta_{\rm{kron-psf}}$ that define galaxy samples of a given completeness, these cuts were measured from the histograms in Fig \ref{fig:sgHist}. The cut defines a minimum $\Delta_{\rm{kron-psf}}$ for stars or a maximum value for galaxies. The dashed lines are fits to the cuts using a second order polynomial of the form 
\begin{equation}
\label{eq:sgCut}
r_{\rm{P1, raw}} - r_{\rm{PSF,raw}} =   \displaystyle\sum\limits_{i=0}^2 a_{i}(r_{\rm{P1, raw}}-21)^{i} .
\end{equation}
We use the 98\% cut to define galaxies throughout this work. Table \ref{table:coeff} gives the values of the coefficients of this equation for different samples. For our adopted cut we again use our synthetic objects, along with fits to the observed SAS2 bright star and galaxy number counts (shown in Fig.~\ref{fig:numberCountsCorr}), to predict completeness and stellar contamination rates. In Fig.~\ref{fig:sgCompl} the predicted galaxy completeness line follows the 98\% line (solid, black), by construction, down to a faint magnitude limit. Near the end of this magnitude range the completeness does drop very slightly and this suggests our fits with Eq. \ref{eq:sgCut} cannot be used beyond a faint magnitude limit of $r_{\rm{P1, kron}} = 23.0$. The dotted line in Fig.~\ref{fig:sgCompl} shows the completeness of the sample after applying the extreme $\Delta_{\rm{kron-psf}}$. This cut, again by construction, has very little effect on the completeness of real galaxies. 

\begin{figure*}
  \includegraphics[width=7.0in]{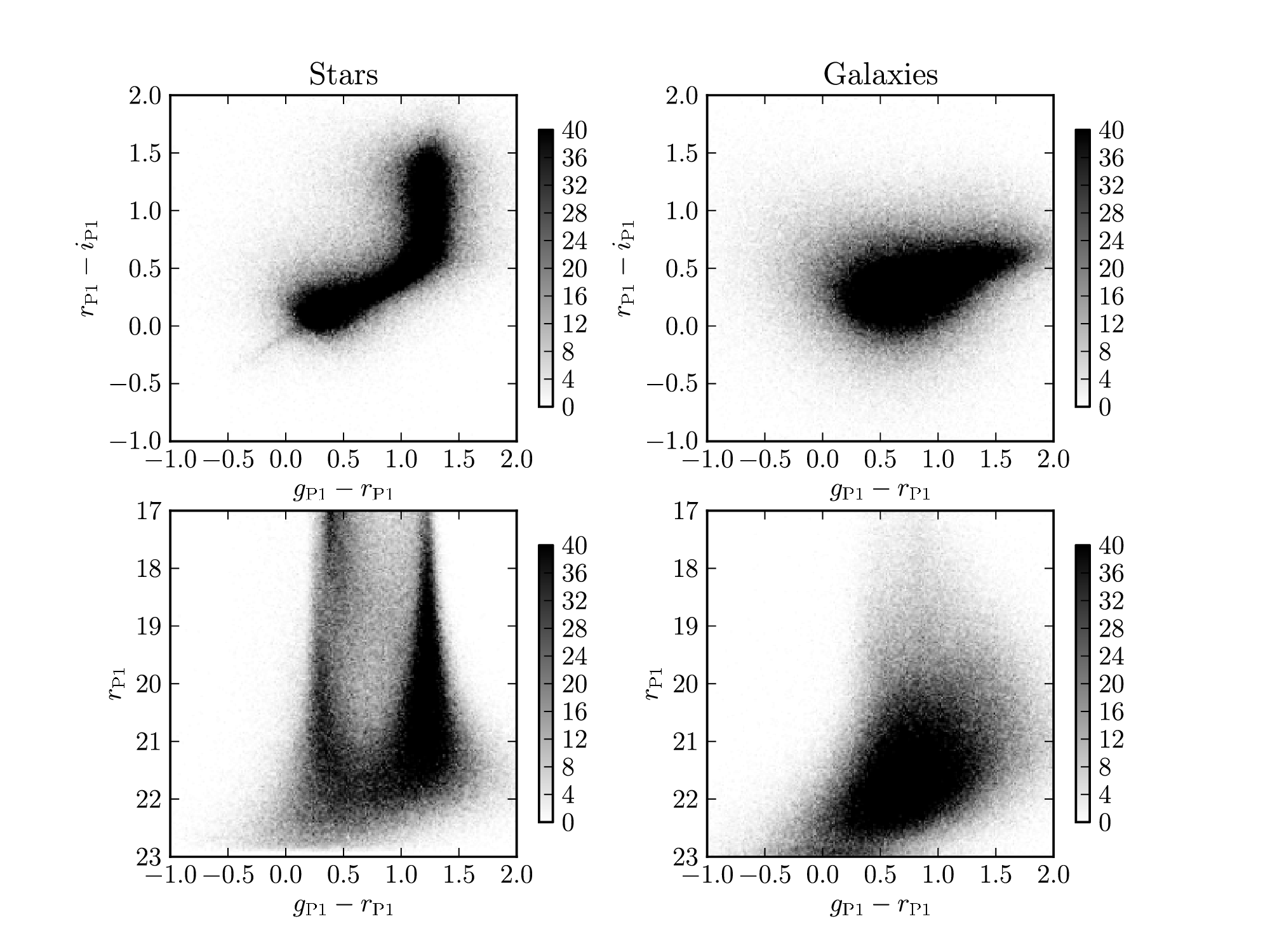}
  \caption{Colour-colour and colour-magnitude diagrams, using Kron magnitudes, of SAS2 objects falling on the star side and galaxy side of our chosen star/galaxy separator (section \ref{sec:sgSep}). The greyscale bar gives the number of objects in each colour-magnitude bin. We see the characteristic stellar features highlighted in Finlator et al (2000), such as the upturn in the colour-colour diagram. In support of our classification we see no evidence of these features in the galaxy sample. }\label{fig:colour}
\end{figure*}

Fig.~\ref{fig:sgCompl} also gives the probability of misclassifying a star as a galaxy (solid red) and the predicted stellar contamination as a fraction of the galaxy sample (dashed red). The latter were calculated from our power law fits to the observed SAS2 bright star and galaxy number counts (Fig.~\ref{fig:numberCountsCorr}). We see stellar contamination stays below 10\% for all magnitude ranges.       
\begin{table*}
\caption{Coefficients for the star, galaxy and false positives separator. Percentages represent the percentage of objects would be included in the sample. The upper or lower limit column defines the direction of the cut, e.g. an upper limit indicates only taking values below the given $r_{\rm{P1,raw}} - r_{\rm{P1,PSF,raw}}$ line.}
\begin{tabular}{c c c c c c}
\hline\hline
Sample & $a_{2}$ & $a_{1}$ & $a_{0}$ & Upper or Lower Limit\\
\hline
98\% Galaxies & $0.018$ & $0.120$ & $-0.192$ & Upper\\
95\% Galaxies & $0.014$ & $0.129$ & $-0.261$ & Upper \\
90\% Galaxies & $0.007$ & $0.129$ & $-0.319$ & Upper \\
Extreme $\Delta_{\rm{kron-psf}}$ & $-$  & $0.417$ & $-1.759$ & Lower \\
\hline
\end{tabular}\label{table:coeff}
\end{table*}
In order to further test our star/galaxy separator, we match our \rps-band data to the \ips and \gps-bands and plot the colour-colour and colour-magnitude diagrams for stars and galaxies classified via our 98\% cut in the \rps-band. The diagrams in Fig.~\ref{fig:colour} follow those for SDSS objects seen in \cite{finlator}. In \cite{finlator} the shape of the distribution of stars in these plots is explained as being driven by different spectral types, with M dwarfs causing the upturn in the colour-colour diagram and F and G disc stars along with fainter, bluer halo stars causing the locus at $\gps-\rps \sim 0.4$. We see no evidence of these features in objects classified as galaxies, which gives further support to the effectiveness of our star and galaxy separator.

\subsection{Comparison to VVDS Spectroscopic Star and Galaxy Classification}\label{sec:vvds}
As a further test of our star/galaxy separator, we compare to the spectral classifications from the F22 field VIMOS VLT Deep Survey (VVDS) \citep{fevre05}, which we downloaded from the CeSAM website\footnote[3]{http://www.lam.fr/cesam/?lang=en}. The VVDS survey is an $I_{\rm{AB}}$ selected sample of objects. Objects targeted for redshifts are purely selected on apparent $I_{\rm{AB}}$ magnitude to be $17.5<I_{\rm{AB}}<22.5$, though the full photometric catalogue is deeper than PS1 \citep{mccracken03, fevre05}. F22 and SAS2 overlap by $~4$ square degrees. We match the two catalogues using a $1\arcsec$ matching radius. From the matched catalogue we select objects which have been targeted for spectroscopy based on the value of the column  {\sc zflags}, taking {\sc zflags=99} to mean the object was not targeted. Following \cite{ilbert05} we also use {\sc zflags} to select objects with secure redshifts, by requiring the last digit of {\sc zflags} to be greater than or equal to 2. Objects with these {\sc zflags} are expected to have the correct redshift 80-99\% of the time, depending on their value of {\sc zflags} \citep{fevre05}. 

\begin{figure}
 \includegraphics[width=3.6in]{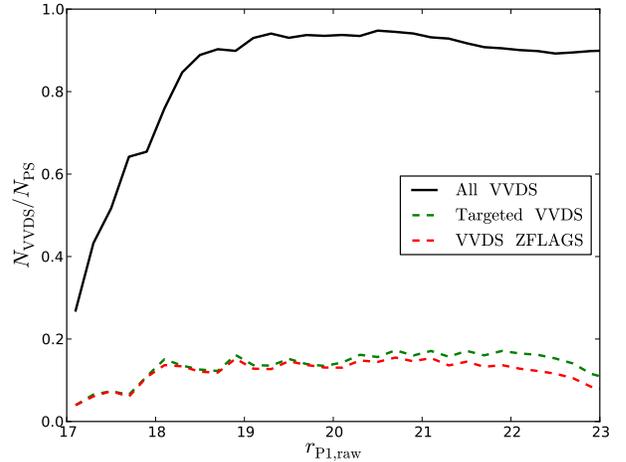}
 \caption{The fraction of Pan-STARRS objects in VVDS as a function of PS1 magnitude: for all VVDS sources (black), VVDS sources targeted for spectroscopy (green-dashed) and VVDS sources with good redshift flags as described in the text (red dashed).}\label{fig:vvdsComplete}
\end{figure}

In Fig.~\ref{fig:vvdsComplete} we show the fraction of objects in PS1 matched to VVDS as a function of PS1 raw Kron magnitude. We do not correct for the VVDS mask, which explains why the curve does not reach unity. An $I_{\rm{AB}}$-band selected sample may have a different morphological mix than an $r$-band selected sample in the same magnitude range. From Fig.~\ref{fig:vvdsComplete} we see the fraction of objects targeted for spectroscopy drops brighter than around $r_{\rm{P1,raw}}>18.0$ and fainter than $r_{\rm{P1,raw}}<22.0$: this is the region where the effects of the VVDS  $I_{\rm{AB}}$-band selection may become important and as such results from these magnitude ranges may be unreliable. Also note from Fig.~\ref{fig:vvdsComplete} the fraction of objects with secure redshifts decreases with magnitude, as one might expect.

A well reported issue in VVDS is its bias against extended sources. Whilst the targeting criteria is purely based on apparent magnitude the program which allocates VIMOS 
slits to targets, the Slit Positioning Optimization Code (SPOC) \citep{bottini05}, is biased against extended sources as they take up more space on the $x$-axis of the 
spectrograph and so decrease the efficiency with which spectra are taken \citep{bottini05}. When computing luminosity functions \citet{ilbert05} corrected for 
this incompleteness by weighting galaxies in a way proportional to their $x$-axis size on VIMOS. We choose to weight galaxies depending on their $\Delta_{\rm{kron-psf}}$. In 
magnitude and $\Delta_{\rm{kron-psf}}$ bins we measure the completeness as the ratio of objects with good {\sc zflags} to all objects matched between PS1 and VVDS 
in the overlap region. The weight of each object is then the inverse of the completeness of its magnitude and $\Delta_{\rm{kron-psf}}$ bin. 

When comparing to VVDS there are three different cases to consider. The first case is where the object is classed as a galaxy in VVDS and PS1, 
we label weights for these objects as $W_{\rm{gg}}$. The second case is for an object classed as a galaxy in PS1 but has a VVDS stellar spectral 
classification, we label weights for these objects as $W_{\rm{gs}}$. The final case is an object classed as a star in PS1 but with a galaxy spectra in VVDS, 
these objects are assigned weights labelled $W_{\rm{sg}}$. The completeness, $C_{\rm{m}}$, and contamination, $C_{\rm{n}}$, 
are estimated using the following weighted sums
\begin{equation}
C_{\rm{m}} = \frac{\Sigma W_{\rm{gg}}}{\Sigma W_{\rm{gg}} + \Sigma W_{\rm{sg}}}, C_{\rm{n}} = \frac{\Sigma W_{\rm{gs}}}{\Sigma W_{\rm{gg}}}  .
\end{equation} 
We plot these estimates, along with jack-knife errors from 9 re-samplings of the data, in Fig.~\ref{fig:sgCompl}. Estimates of stellar contamination are slightly higher than the estimates based on synthetic images, but this is only a small discrepancy given the size of the errors. Estimates of completeness agree until around $\rps=20$ when it looks like our synthetic source estimates are too optimistic. The spectroscopic estimates suggest a completeness of around 91\%, as 
opposed to the predicted 98\%. 

There are several possible reasons for this difference. A major cause of disagreement is likely to be misclassifications in the VVDS sample. To calculate
the fraction of objects in our sample which could be misclassified by VVDS we use {\sc zflags}. The value of {\sc zflags} has been related to the
probability of having been assigned the correct redshift, $P_{\rm{correct}}$, by \citet{fevre05}. We assume this is also the probability of being correctly
classified as a star or galaxy. Table \ref{table:vvdsCompl} gives the different fractions of the full sample, $F_{\rm{Full}}$, and sample with discrepant
star/galaxy classification,  $F_{\rm{Disagree}}$, that have certain values of {\sc zflags}. Table \ref{table:vvdsCompl} also gives our estimate of the 
fraction of objects in the full sample with incorrect VVDS classification, $F_{\rm{Full}}(1-P_{\rm{correct}})$. Given that 9\% of the full sample have
discrepant classfications, $0.09F_{\rm{Disagree}}$ is the number of objects in the discrepant sample with a certain {\sc zflags} value as a fraction of the total 
matched sample. Taking the minimum of $0.09F_{\rm{Disagree}}$ or $F_{\rm{Full}}(1-P_{\rm{correct}})$ for each {\sc zflags} value in Table \ref{table:vvdsCompl} and 
summing suggests that 6\% of our disagreement could be down to misclassified VVDS objects. This would lead to a VVDS misclassification-corrected estimate of
completeness of 97\%, consistent within random errors with our estimate from synthetic sources.  

\begin{table}
\begin{center}
\caption{The second column gives the probability of a redshift measurement being correct for different {\sc zflags} values, taken from \citet{fevre05}. 
The third and forth columns give the fraction of the full sample and discrepant sample that have certain {\sc zflags} values. The final column gives an
estimate of the total fraction of PS1 objects with incorrect VVDS estimates of redshift.} 
\begin{tabular}{c c c c c}
\hline\hline
{\sc zflags} & $P_{\rm{correct}}$ & $F_{\rm{Full}}$ & $F_{\rm{Disagree}}$ & $F_{\rm{Full}}(1-P_{\rm{correct}})$ \\
\hline
2 & 0.80 & 0.25 & 0.42 & 0.05 \\
3 & 0.91 & 0.22 & 0.23 & 0.02 \\
\hline
\end{tabular}\label{table:vvdsCompl}
\end{center}
\end{table}

Another potential explanation is that the synthetic galaxies may be slightly too extended in their $\Delta_{\rm{kron-psf}}$ values. Simplification of modelling galaxies with de Vaucouleurs and exponential profiles, adopting a mean extinction value for the galaxies, using redshifts and magnitudes from {\sc galform} and 
only generating synthetic images on one skycell could all contribute to this effect.

From Fig.~\ref{fig:sgCompl} it appears that our classification is around 91\%-98\% accurate down to faint magnitudes depending on how you estimate classification completeness. Brighter than $r_{\rm{P1,raw}}=22.0$ stellar contamination is below 6\%, increasing to around 10\% at magnitudes fainter than this. The action of stellar contamination, on smaller scales where the stars are uniformly distributed, is to dilute the clustering by $(1-f)^2$, where $f$ is the fraction of stars in the galaxy sample \cite[e.g.][]{hudon96, roche99}. We will revisit the effect of stellar contamination in Section~\ref{sec:starFalse}.

Classification contamination and completeness can influence galaxy clustering measurements and as such work on star and galaxy separation is ongoing. Classifications based on SED fits along with star/galaxy separators calibrated on other data sets and other morphological measurements will be available to help meet the future PS1 science goals.

\section[Variable Depth]{Dealing with Variable Depth}\label{sec:detectionEff}
The finished PS1 3$\pi$ survey will have spatially variable image depth for several reasons. These include spatially varying stack coverage due to masking and greater or fewer visits to any piece of sky (see Fig.~\ref{fig:cov}), varying PSFs and varying sky brightness. To measure reliable clustering it is vital to measure the angular incompleteness, otherwise fluctuations in galaxy density caused by changes in depth would contaminate the clustering measurements. Once this angular incompleteness is modelled we can deal with it by introducing the same depth variations into the random distribution of points we use to measure clustering, which we shall refer to from now on as our ``random catalogue''.

We assume that the probability of detecting an object is only dependent on the signal-to-noise ratio. In order to make a simplified estimate of the signal to noise ratio we assume all sources have a Gaussian light distribution. For the stacked data most galaxies near the magnitude limit have small angular sizes so this is a reasonable approximation (we further test this later in this section). Using a PS1 PSF rather than a Gaussian would simply scale our FWHM measurements to different values, an effect that would be removed by the empirical calibration we present later in this section. We define the ``fiducial'' SNR as 

\begin{equation}
\label{eq:fidSNR}
\mathrm{SNR} = \frac{F}{\sqrt{\pi d_{\mathrm{FWHM}}^2 \sigma^2}},
\end{equation}
where $d_{\rm{FWHM}}$ is the FWHM of the PSF in units of pixels, $F$ is the apparent flux of the source (without extinction correction) and $\sigma^2$ is the variance according to the variance map. Whilst $d_{\rm{FWHM}}$ is measured for all PS1 detections, for this work we use the typical FWHM of SAS2 of $0.94\arcsec$. As SAS2 has fairly uniform seeing this simplifies our work whilst not affecting our results. We use our masks to extract $\sigma^2$ which results in the loss of some spatial accuracy. This is unavoidable due to the otherwise prohibitively slow process of retrieving the individual variance maps at the native pixel scale.
\begin{figure*}
\begin{center}
 \includegraphics[width=6.5in]{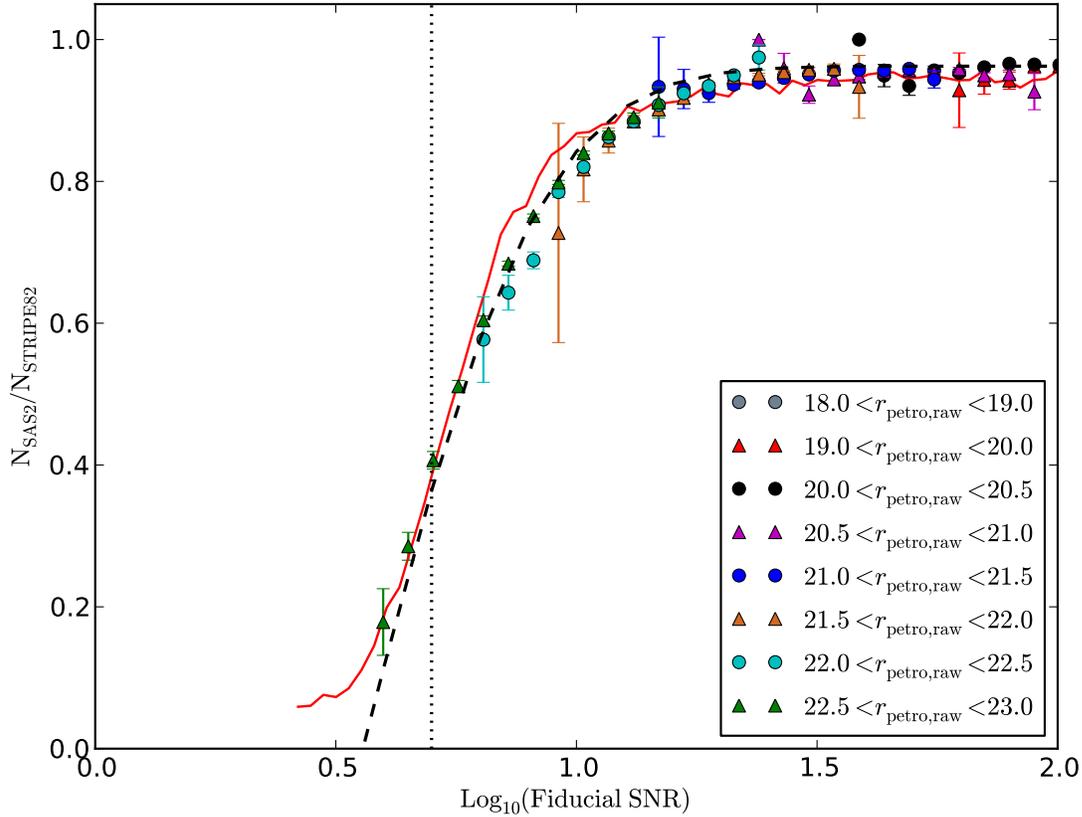}
 \caption{The fraction of Stripe 82 objects detected as a function of fiducial SNR (Eq. \ref{eq:fidSNR}). Overlap between magnitude bins implies the fiducial SNR can be used as an estimator of the probability of source detection. The red line shows this quantity as measured from the synthetic galaxies added into real PS1 images and processed by the standard IPP. The dashed line shows the best-fitting relation of Eq. \ref{eq:pDet}, the dotted line marks $\rm{SNR}=5.0$. Error bars are from 100 bootstrap re-samplings.}\label{fig:snr}
\end{center}
\end{figure*}

To calibrate the relationship between our fiducial SNR measurements and source recovery fraction we again make use of the overlap region with Stripe 82. We use the Stripe 82 Petrosian magnitude to calculate the fiducial SNR for all Stripe 82 galaxies and then match to PS1 SAS2 and see what fraction are recovered. We plot these fractions in Fig.~\ref{fig:snr} in different magnitude bins. The fact that over different magnitude bins the fiducial SNR values have the same detected fraction shows that this measurement can be used to assess the probability of detection. We parameterize this curve with the fitting formula

\begin{equation}
\label{eq:pDet}
P_{\mathrm{Det}}(\mathrm{SNR}) = a\;\mathrm{erf}(b\;\mathrm{log_{10}}(\mathrm{SNR}) + c),
\end{equation}
where $a$, $b$ and $c$ are constants with best-fitting values $a=0.962$, $b=2.446$ and $c=-1.361$. The fact that $a$ is not unity implies there is always some fraction of Stripe 82 objects undetected by PS1. We believe this fraction is caused by false positives in SDSS Stripe 82 and visually inspecting a subset of these objects suggests they are mainly caused by spurious detections in the wings of extended objects. So long as the number of false positives in Stripe 82 remains a constant fraction of the real objects this effect should not bias our results, this seems to be the case as the curve is flat for large values of fiducial SNR. As a sanity check we also add a curve to Fig.~\ref{fig:snr} showing the detection efficiency estimated from our synthetic galaxies, using the input synthetic object magnitude corrected to Kron magnitude using a correction of 0.2 magnitudes (explained in Section \ref{sec:data}). Our estimate of detection efficiency from synthetic objects shows a reasonable agreement with the real data on the plot, though the synthetic galaxies seem to suggest the Stripe 82 comparisons slightly underestimate the depth at fiducial SNR values of around 6 to 9. The differences could be due to multiple causes. For example, it could be Stripe 82 false positives or slightly above average seeing in the skycell used in Section \ref{sec:sgSep}. As these differences are only of the order of a few percent we choose to defer further careful studies to the analysis of the full $3\pi$ dataset, where a larger amount of deeper comparison data will be available.      

We can see in Fig.~\ref{fig:snr} that a $5\sigma$ SNR implies a 20-30\% detected fraction which is lower than the measurements in Paper I of 50-60\% recovery of fake stars at that SNR, this is to be expected as extended objects at these magnitudes are lower surface brightness and therefore harder to detect. None the less this highlights the fact that where the curve of Fig.~\ref{fig:snr} is steep small changes in the SNR can lead to large changes in detection fraction. To avoid any problems caused by this we impose a default lower limit on the fiducial SNR by excluding spatial regions where $\rm{SNR}<3.0$ . We experiment with different values of this parameter in Section \ref{sec:clustering}. 

\begin{figure*}
\begin{center}
\includegraphics[width=6.0in]{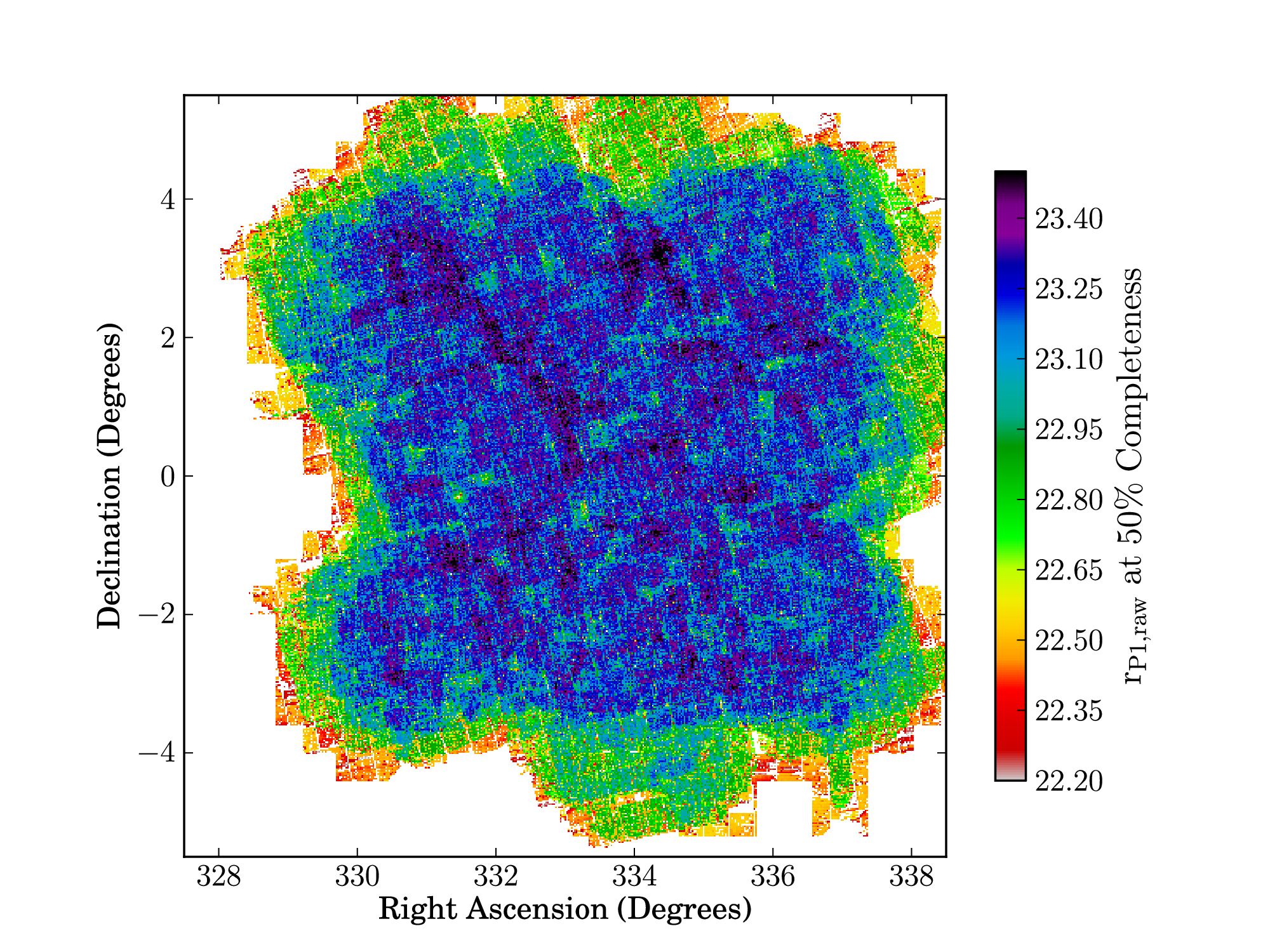}
 \caption{The $\rm{r_{P1,raw}}$ magnitude corresponding to 50\% galaxy completeness as predicted by our fiducial SNR method. SAS2 is shallower near the edges where there are fewer exposures, while the pattern of deeper areas across the central region is more representative of of what we expect from the whole 3$\pi$ survey.}\label{fig:spatialDepthVariation}
\end{center}
\end{figure*}

Using our binned up variance maps and Eq.~\ref{eq:pDet} we can produce a map of the magnitude at 50\% galaxy recovery, shown in Fig.~\ref{fig:spatialDepthVariation}. Note we can produce these maps even in SAS2 regions without Stripe 82 overlap, as we only need Stripe 82 to calibrate Eq.~\ref{eq:pDet}. One can clearly see the shallower regions near the edges of the SAS2 field, along with patterns of deeper regions in the central area caused by the overlapping pattern of input exposures. Fig.~\ref{fig:spatialDepthVariation} demonstrates our technique produces maps of depth to very high resolution, contrast this with the much lower resolution depth maps produced using synthetic stars presented in figure 15 of Paper I. Reassuringly we see common features, including the shallower edge region and the deeper diagonal feature.   

We use the curve measured in Fig.~\ref{fig:snr} to correct our random catalogue by making the chance of placing a random point of a certain magnitude in any region equal to the detected fraction expected for that region given the random point's fiducial SNR. Magnitudes are assigned to the random points from the observed galaxy counts, uncorrected for extinction. As a first pass we estimate these number counts by fitting the bright end of the galaxy counts with a power law (in Section \ref{sec:ncounts} we show we can use our method to yield depth corrected number counts, which we use to assign magnitudes to the random points). After assigning magnitudes and deciding if a random is detected, we extinction-correct the random catalogue. This technique results in a random catalogue with the same spatial depth variation as the data. 

\begin{figure*}
\begin{center}
\includegraphics[width=6.0in]{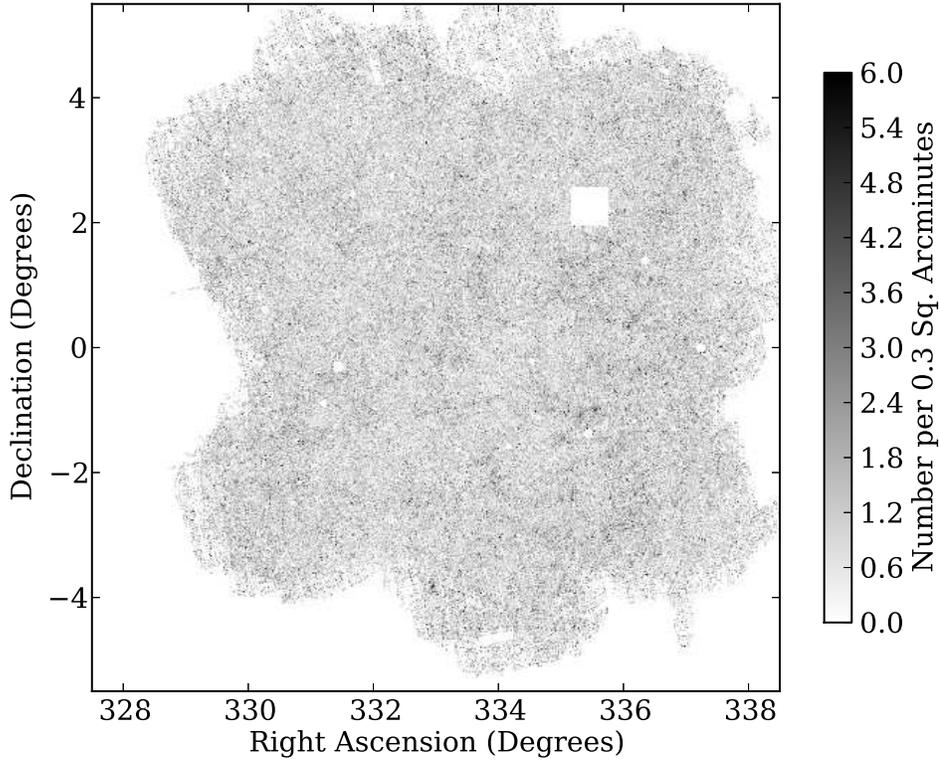}
 \caption{The number density of galaxies, binned by right ascension and declination and corrected for variable depth. We claim over-densities in this plot are genuine, except those caused by Poisson noise in pixels nearer the edges of the field which have small numbers of galaxies.}\label{fig:correctedNumberDensity}
\end{center}
\end{figure*}

We plot the depth-corrected density of galaxies in Fig.~\ref{fig:correctedNumberDensity}. To produce this figure we binned the galaxies and detection efficiency randoms onto the same grid and then divided the galaxy grid by the random grid, normalising by the ratio of the relative numbers of galaxies and randoms. To eliminate noise from regions with very few randoms, generally near the edge of the field, we white-out pixels with fewer than 5 randoms. Comparing Fig.~\ref{fig:mask}(bottom) to Fig.~\ref{fig:correctedNumberDensity} we see the over-densities caused by varying image depth are removed. There are fewer objects in Fig.~\ref{fig:correctedNumberDensity} than Fig.~\ref{fig:mask}(bottom) as star/galaxy separation has removed the stars.

\begin{figure}
\begin{center}
\includegraphics[width=3.3in]{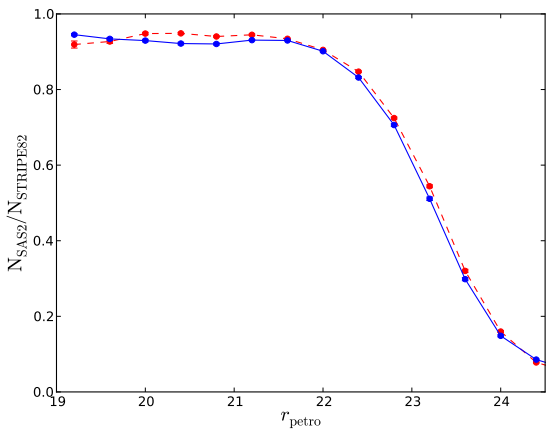}
 \caption{The detected fraction of Stripe 82 galaxies, separated into red and blue by Stripe 82 colours. We see no evidence that red and blue galaxies have different detection efficiency properties, despite the fact that their morphology is expected to be different.}\label{fig:colourVDepth}
\end{center}
\end{figure}

One key assumption of our depth correction method is that all galaxies in our sample have the same detection efficiency properties for the same fiducial SNR, i.e. that secondary parameters such as morphology or colour are unimportant in determining how likely objects are to be detected (consider Eq. \ref{eq:fidSNR}). We argue that for faint magnitudes galaxies predominantly have small angular sizes and as such look similar to one another after being convolved with the PSF. To further test this we plot, in Fig.~\ref{fig:colourVDepth}, the detection efficiency curves of our Stripe 82 red and blue samples of galaxies. In Fig.~\ref{fig:colourVDepth} we see, for the same reasons as in Fig.~\ref{fig:snr}, that the curve does not reach unity. We also see that at brighter magnitudes blue galaxies have a lower detection efficiency. As this effect is at magnitudes far brighter than our detection limit we attribute this to false positives in Stripe 82 falling on the blue side of our colour cut. The agreement between the red and blue detection efficiency curve at faint magnitudes in Fig.~\ref{fig:colourVDepth} suggests that an undetected low surface brightness population of galaxies must either be split equally between our two colour bins or represent a very small fraction of our sample. This supports our assumption that at the limiting magnitude of 3$\pi$ data detection efficiency depends on a single parameter, SNR. However in small regions of the 3$\pi$ survey where the limiting magnitude may be much brighter, and galaxies near this magnitude have larger angular sizes, the situation may be more complicated. 

Another important thing to note is that our method gives a measurement of the clustering of only the detected galaxies. As our faintest samples
become incomplete towards faint magnitudes they will, to some extent, be biased towards bright galaxies. To test how much of an effect
this is we utilize our random catalogues, measuring the median magnitude of random samples before and after degrading them for detection efficiency effects.
In this paper we mainly use 0.5 magnitude bins, here the difference in median magnitudes is less than 0.04 magnitudes for the faintest sample of 
$22.5<\rps<23.0$, and zero for samples brighter than around  $\rps = 21.5$. Our measurements of clustering in these bins will not be significantly affected by this
small difference. In general, however, it is important to note that this method assumes the detected galaxies have the clustering properties of the full population, i.e. 
the method only corrects for the spatial dependence of image depth. Care will have to be taken to not apply this method where there is a large 
variation in completeness across the apparent magnitude range of a sample.

\section[Angular Clustering]{Results and Tests for Systematics}
\subsection{Number Counts}\label{sec:ncounts}
\begin{figure}
\begin{center}
\includegraphics[width=3.3in]{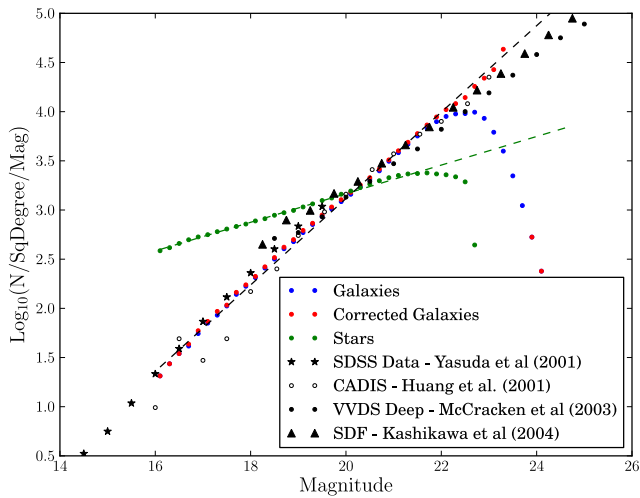}
 \caption{Number counts in the \rps-band before (blue) and after (red) the depth correction for galaxies, along with the number counts of objects classed as stars by our adopted separator (green). We do not correct the stars, or the galaxies fainter than $r_{P1}=23.7$ for completeness. The dashed lines are power law fits to the number counts. Example $r$-band literature galaxy counts have been included, as indicated in the legend. PS1 Kron magnitudes have been corrected to total using our adopted correction of 0.2 magnitudes.}\label{fig:numberCountsCorr}
\end{center}
\end{figure}

We plot, in Fig.~\ref{fig:numberCountsCorr}, the \rps-band differential number magnitude counts of galaxies before and after our correction. A Kron to total correction of 0.2 magnitudes is applied to the galaxy counts, as explained in Section \ref{sec:data}. To generate the detection efficiency corrected number counts in Fig.~\ref{fig:numberCountsCorr} we use our extinction corrected random catalogue, from before and after the detection efficiency corrections, to predict the fraction of galaxies detected as a function of extinction corrected magnitude. We then correct the observed number counts by these fractions. We see after the counts have been corrected the turnover no longer occurs, and the counts continue to grow to very faint magnitudes until we stop using our depth correction at $r_{P1}=23.7$, where the correction is very large (a factor of 70 at this magnitude). We see that our number counts show reasonable agreement with the published data of \citet{huang01}, \citet{yasuda2001}, \citet{mccracken03} and \citet{kashikawa04}. At the faintest magnitude our number counts are slightly above the literature measurements, both where our correction is and is not important. This could be partially due to the 10\% false positives at these magnitudes (see Fig.~\ref{fig:hist} and also Paper I). It could also be partially explained by cosmic variance, as the literature measurements also disagree to a similar extent at these magnitudes.

\begin{figure*}
\begin{center}
 \includegraphics[width=6.5in]{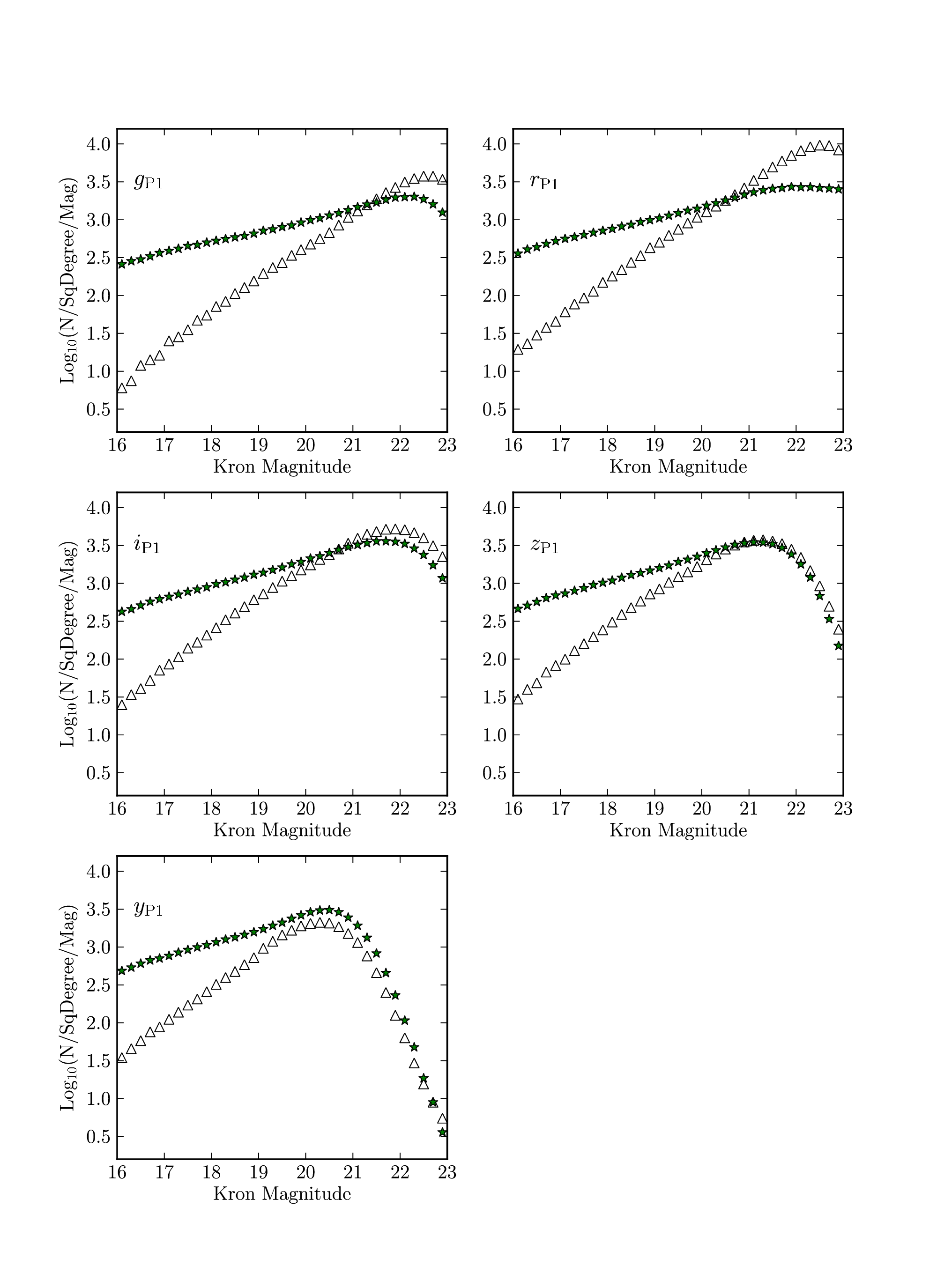}
 \caption{The differential number magnitude counts for stars (green stars) and galaxies (open triangles) for different PS1 bands matched to the \rps-band, in which the star and galaxy separation cut was applied (section \ref{sec:sgSep}). The matching only limits the depth of the \ips-band, as this band is the only one deeper than the \rps-band.}\label{fig:ncounts}
\end{center}
\end{figure*}

In Fig.~\ref{fig:ncounts} we show our measured, uncorrected number counts for different bands. Each band was matched to the \rps-band, where the star/galaxy classification was made. The galaxies show a power law trend in good agreement with previous measurements. The stars show a shallower power law trend. The turnover in the samples is caused by the incompleteness and this turnover happens at brighter magnitudes as we move toward redder bands. In redder bands the ratio of stars to galaxies increases, until the \yps-band where we see more stars than galaxies at all magnitudes. As these are the same objects as seen in Fig.~\ref{fig:numberCountsCorr} the main purpose of this plot is to check if our \rps-band star/galaxy classification gives sensible results for different bands. We leave detailed science analyses using the number counts to later work. 

\subsection{Angular Clustering}\label{sec:clustering}
In this section we present measurements of angular clustering. To measure this clustering we make use of the GPU code of \cite{bard}. We use the \cite{hamilton} estimator, though our results are unchanged if we use the \cite{landySzalay} estimator. Error bars for all clustering measurements are from 9 jack-knife re-samplings of the data. We use eight times as many random points as data points throughout. On each clustering plot we draw the same dashed-black reference line, for easier comparisons between plots. 

When measuring clustering an effect known as the integral constraint can artificially weaken clustering on scales comparable to the area of the survey \citep[e.g.][]{roche99}. For  SAS2 data, over the scales we measure clustering, this has no effect on our results, except in one case we will discuss later. For the MD07 measurements however the smaller area results in the integral constraint being important on the scales we consider. We therefore estimate the true clustering of the MD07 data on large scales by fitting a power law between scales of 0.002 to 0.165 degrees and then use this fit to estimate the size of the integral constraint using the standard formula \citep[e.g. equation 9 of][]{roche99}. As an example, for our threshold sample $\rps<23.0$ the integral constraint is 80\% of the signal at the largest separations plotted, dropping to 14\% by $\theta \approx 0.1$ deg.

\begin{figure}
\begin{center}
\includegraphics[width=3.6in]{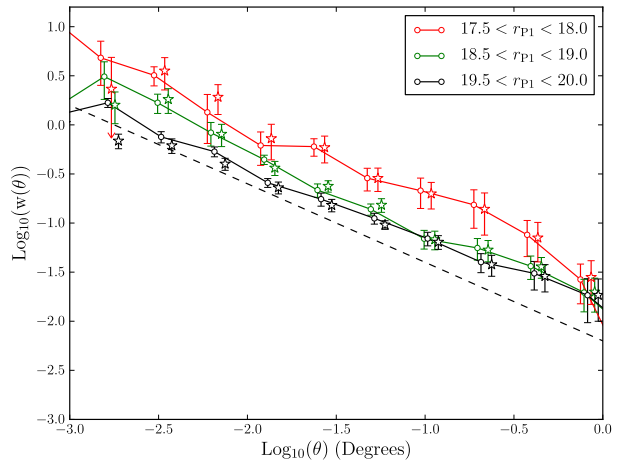}
 \caption{Angular clustering of galaxies in PS1 (connected, open circles) and in the same region of SDSS DR8 (star-shaped symbols), both measured for this paper using the sample selection described in the text. This shows good agreement between PS1 and SDSS DR8. The dashed black line is a reference line included in all of our clustering plots. Different measurements have been offset horizontally for clarity, the brightest galaxies are at the true $x$-axis position for all of the measurements.}\label{fig:clusteringBright}
\end{center}
\end{figure}

We begin by studying the regime where the spatially varying depth correction has no effect. Fig.~\ref{fig:clusteringBright} shows the clustering of PS1 data compared the the clustering of DR8 data over the same region, which we measured from our galaxy sample (Section \ref{sec:sdss}). We see the well-reported effect of clustering being stronger in brighter apparent magnitude bins. This result is caused by two effects. The first is that fainter magnitude bins are projected over larger radial distance ranges so incoherent clustering signals are summed together decreasing the clustering strength. The second cause is that intrinsically fainter galaxies are less clustered, usually interpreted as evidence they lie in less massive dark matter haloes. This latter effect is much smaller than the former as apparent magnitude ranges relate to similar absolute magnitude ranges. We see good agreement between the SDSS and PS1 measurements for these ranges, an agreement much closer than the jack-knife error bars as the two data samples are from the same area of sky. We do see some differences, but photometric errors scatter galaxies in and out of the different magnitude bins and so the two samples can contain a significant fraction of galaxies that are not in common. Overall, Fig.~\ref{fig:clusteringBright} acts as a detailed test to determine if PS1 is capable of measuring the clustering of galaxies down to $\rps=20.0$. Fainter than this it becomes more difficult to measure reliable clustering with SDSS DR8 and as such we compare to measurements in the literature.

\begin{figure}
\begin{center}
\includegraphics[width=3.6in]{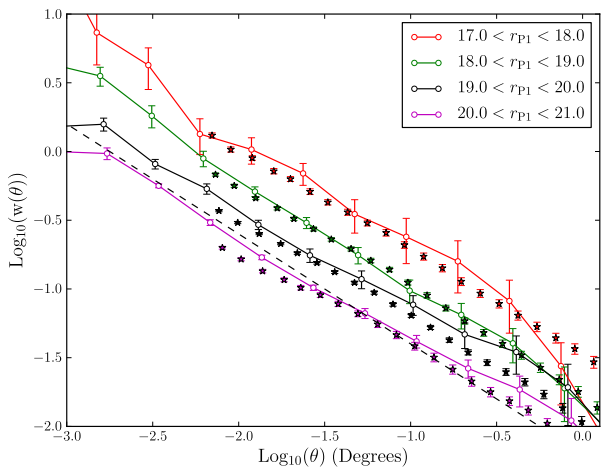}
 \caption{Angular clustering of galaxies over the full SAS2 area in PS1 (connected, open circles) and measurements from a much larger area of SDSS DR7 (filled stars) from Wang et al. 2013. The dashed black line is the same reference power law as in Fig.~19. Different measurements have been offset horizontally for clarity, the brightest galaxies are at the true $x$-axis position for all of the measurements. Different measurements have been offset horizontally for clarity, with the brightest SDSS and PS1 samples showing the position of the true angular bins.}\label{fig:clusteringWang}
\end{center}
\end{figure}

In Fig.~\ref{fig:clusteringWang} we compare our angular clustering measurements from PS1 SAS2 to recent angular clustering measurements from \cite{wang13} from 8000 square degrees of SDSS DR7 data. In \cite{wang13} careful studies are carried out which suggest SDSS DR7 can measure clustering down to $r=21.0$,  Fig.~\ref{fig:clusteringWang} demonstrates PS1 data shows reasonable agreement with the SDSS data. Naturally, differences arise due to sample variance in the relatively small SAS2 field, but Fig.~\ref{fig:clusteringWang} is a promising indicator that PS1 clustering measurements are capable of matching SDSS depth. Fainter than $\rps=21.0$ the spatially varying depth will start to become important. 

\begin{figure}
\begin{center}
\includegraphics[width=3.6in]{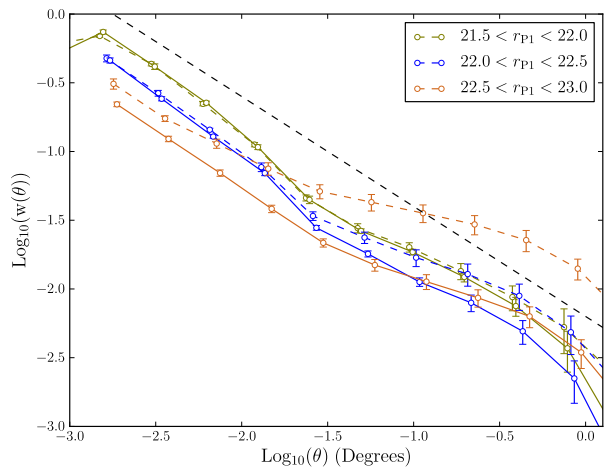}
 \caption{Angular clustering of faint galaxies before (dashed lines with points) and after (solid lines with points) applying our spatially varying depth correction. The dashed black line is the same reference power law as Fig.~19. Different measurements have been offset horizontally for clarity. The uncorrected clustering of the brightest galaxy sample is at the true $x$-axis position for all of the measurements.}\label{fig:clusteringFaint}
\end{center}
\end{figure}

To see the effects of our depth correction we plot, in Fig.~\ref{fig:clusteringFaint}, the 2-point angular correlation function of galaxies before and after correcting the random catalogue for spatially varying depth. At $\rps=22.0$ and $\rps=23.0$, the edges of the brightest and faintest bins in Fig.~\ref{fig:clusteringFaint}, the average completeness is only $80\%$ and $50\%$ respectively. We see that without corrections clustering in these faint bins is enhanced by under-densities and over-densities caused by the spatially varying incompleteness. After correction the clustering strength is decreased, with the effect being more marked for the fainter bins where one would expect the depth to be most spatially inhomogeneous. The strength of clustering in the fainter bins has its largest correction at large scales. Magnitude ranges brighter than $\rps<22.0$ seem to need very little correction, whereas the correction becomes larger for fainter bins. The SAS2 region is more uniform than the full $3\pi$ data so the magnitudes at which the spatial depth variation correction becomes important may differ for the full 3$\pi$ survey.

\begin{figure}
\begin{center}
 \includegraphics[width=3.6in]{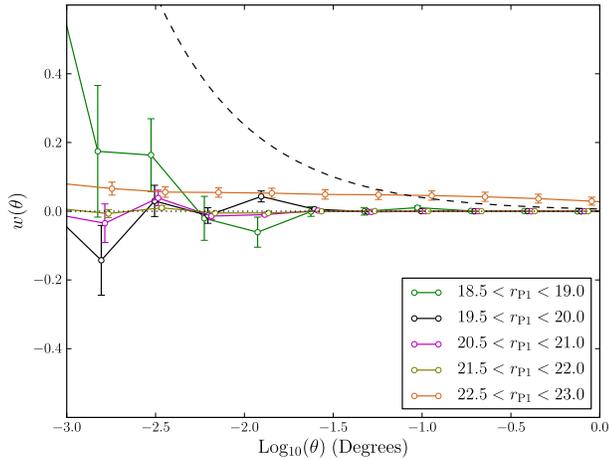}
 \caption{The angular correlation function of a random catalogue that has detection efficiency corrections applied to it, in effect measuring the clustering of the detection-weighted randoms relative to a uniform set of randoms. This gives an estimate of the clustering signal introduced into the data by spatially varying depth. The clustering here is much weaker than the clustering of the galaxies, indeed for most magnitude bins there is no clustering. The faintest magnitude range shows a clear clustering signal, introduced by our modulation of the randoms to correct for spatially varying incompleteness. The dashed black line is a reference power law added to all of our clustering plots, the dotted line marks no clustering.}\label{fig:randClustering}
\end{center}
\end{figure}

As an alternate way of understanding our correction we measure the angular correlation function of our spatial depth corrected random catalogue, relative to an uncorrected, spatially uniform random catalogue. This gives us an estimate of the signal we remove from the faint magnitude bins. We see in Fig.~\ref{fig:randClustering} the clustering of the bright randoms is consistent with no clustering signal. Bright randoms have larger errors as there are fewer bright randoms. For the faintest bin, where we see the strongest correction, the randoms are clustered. This type of clustering signal is the effect of variable depth on our measurements. We can infer that without correction clustering is enhanced on all scales. This effect will be particularly noticeable on larger scales where the intrinsic galaxy clustering is weak, this is seen in Fig.~\ref{fig:clusteringFaint}.

\begin{figure}
\begin{center}
\includegraphics[width=3.6in]{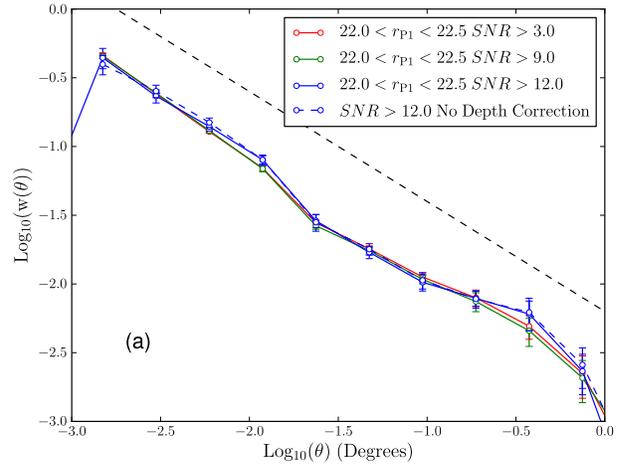}
\includegraphics[width=3.6in]{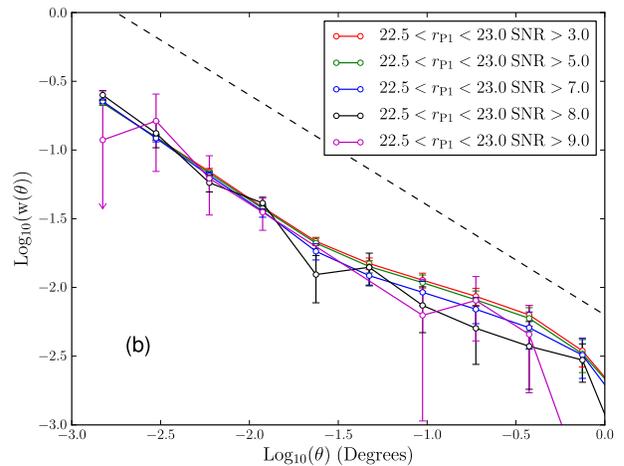}
 \caption{Angular clustering for two different magnitude bins in sub-areas satisfying different fiducial SNR cuts, as indicated in the key. The different panels (a) and (b) show different magnitude ranges. We see that more conservative estimates of the clustering are in agreement with measures which use less deep data with a larger correction applied. For the brighter magnitude bin we also plot the clustering uncorrected for spatially varying depth from a region where the depth is fairly uniform. The points for the different curves have been artificially displaced along the $x$-axis for clearer viewing, the top curve in the legend shows the true $x$-axis position for all curves. The dotted line in panel (b) is the power law we use to roughly estimate the effects of the integral constraint for this panel, which is necessary as the sub-areas in the faintest magnitude bin can be very small. The dashed black line is a reference power law added to all of our clustering plots.}\label{fig:clusteringSNRCuts}
\end{center}
\end{figure}

Qualitatively the correction appears to be doing a good job. To carry out a quantitative test we find the variance value which corresponds to some fiducial SNR at the faint edge of a magnitude bin, and mask spatial regions in the randoms and data that have a variance value higher than this. This limits our depth correction by removing data and randoms with fiducial SNR lower than some limit. The corrected clustering measurements for the range $22.0<\rps<22.5$ in Fig.~\ref{fig:clusteringSNRCuts}(a) are robust to changes in the choice of the SNR limit, with more conservative cuts in SNR being in agreement with the more lenient cuts. To further emphasise this we plot the clustering of $22.0<\rps<22.5$ galaxies with and without spatial depth corrections, in regions with $\rm{SNR}>12.0$ where the depth is fairly uniform. We see from these curves that using the full SAS2 region combined with a correction gives results in agreement with using a smaller region of uniform depth. 

We plot in Fig.~\ref{fig:clusteringSNRCuts}(b) the same tests for the faintest magnitude bin, $22.5<\rps<23.0$. The conservative SNR cuts in the faintest magnitude bin restrict the area of the survey, and as such the integral constraint becomes important. We therefore correct clustering measurements in this plot for the integral constraint, using the power law fit plotted in grey. We do not show cuts more conservative than $\rm{SNR} > 9.0$ as there are very little data beyond that cut in this magnitude range. Unfortunately the results of this test are less convincing, the different $\rm{SNR}$ cuts agree within error but there does appear to be a systematic trend for more conservative $\rm{SNR}$ cuts to measure a slightly weaker clustering signal on larger scales. This could suggest our correction is too small, though it could also be caused by other problems at very faint magnitudes such as false positives. Remember that in this faintest magnitude bin our correction is extremely large and the data is very incomplete. Completeness is only around 50\% at $\rps=23.0$ (Paper I), so it is perhaps not surprising that the method is less successful in this regime. For science applications we do not intend
to apply corrections as large as this. Instead we would place a limit on the minimum SNR of the data analysed and so would exclude the shallower areas of the survey when 
constructing the faintest datasets. However, the fact that our method is reasonably successful in this regime is a positive indication that our method will work for more 
uniform data.

\begin{figure}
\begin{center}
\includegraphics[width=3.6in]{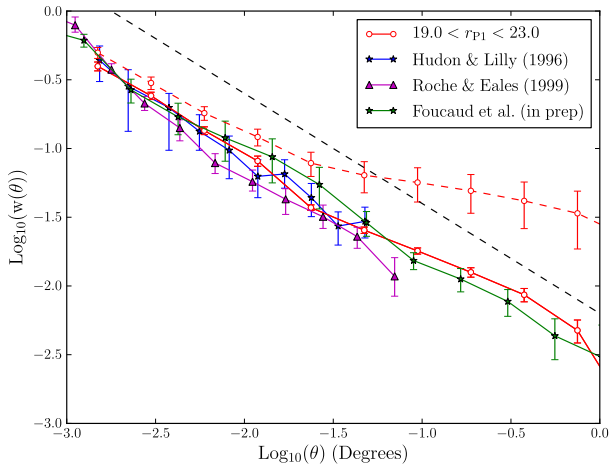}
 \caption{A comparison of our measurements to Hudon \& Lilly (1996), Roche \& Eales (1999) and the PS1 MD07 measurements of Foucaud et al (in preparation), before (dashed) and after (solid) our depth correction. The depth correction brings our results into closer agreement with the other measurements, which are from deeper and more uniform surveys than the PS1 SAS2 data. No attempt has been made to correct for the differences between the Hudon \& Lilly (1996) or Roche \& Eales (1999) $R$-band filters and our \rps-band filter. The dashed line is a reference power law added to all of our clustering plots.}\label{fig:clustering2}
\end{center}
\end{figure}

In Fig.~\ref{fig:clustering2} we compare our measurements of clustering to those of \citet{hudon96}, field ``e'' of \citet{roche99} and Foucaud et al. (in preparation) for the magnitude range $19.0<\rps<23.0$. Note that the \citet{roche99} sample is for $18.5<R<23.0$ measured in the Vega system, but despite these small differences it is still a useful comparison. The amplitudes of \citet{hudon96} and \citet{roche99} have been corrected for stellar contamination using their estimate of the contamination fraction of $f=0.29$ and $f=0.11$ respectively. As introduced in Section~\ref{sec:vvds} this correction is boosting the amplitude by $(1-f)^{-2}$ and is the same correction  \citet{hudon96} and \citet{roche99} apply to their own results. We estimate our contamination fraction, from the dashed red line in Fig.~\ref{fig:sgCompl}, to be $f=0.07$ for this sample and we correct our amplitude accordingly. Foucaud et al. (in preparation) estimate their stellar contamination to be $f=0.06$, so we also correct their clustering measurements. 

In Fig.~\ref{fig:clustering2} we see our depth correction brings us closer to the other measurements of clustering. On smaller scales we show reasonable agreement 
with the literature measurements of \citet{hudon96} and \citet{roche99}. Within the errors we show agreement with the MD07 clustering measurements of 
Foucaud et al (in preparation). Our correlation function is slightly higher than the MD07 measurements on scales greater than $10^{-1.5}$ degrees, and slightly 
lower on scales less than this. However, these differences are within the reasonably large error bars of the MD07 sample. The scatter in the literature measurements 
is also large due to the small size of the samples. As such current available comparison data in the $r$-band is limited by sample variance, limiting our ability 
to assess any remaining systematic errors. Another limitation of this comparison is that, for this magnitude range, the median magnitude of our incomplete sample of 
galaxies will be 0.16 magnitudes brighter than that of a complete sample. As explained, this occurs where the completeness of the galaxy sample shows significant variation
across the sample's apparent magnitude range. This would lead to our measurements having a slightly stronger clustering amplitude than
a complete sample, which could explain some of our disagreement with the MD07 clustering measurements on larger scales. Despite the limitations of this comparison plot, it 
is still impressive that in the regime where the corrections are very large our method does a qualitatively good job at recovering the clustering signal. We do not intend our 
method to be applied to such incomplete data for science applications.

\begin{figure*}
\begin{center}
\includegraphics[width=0.9\textwidth]{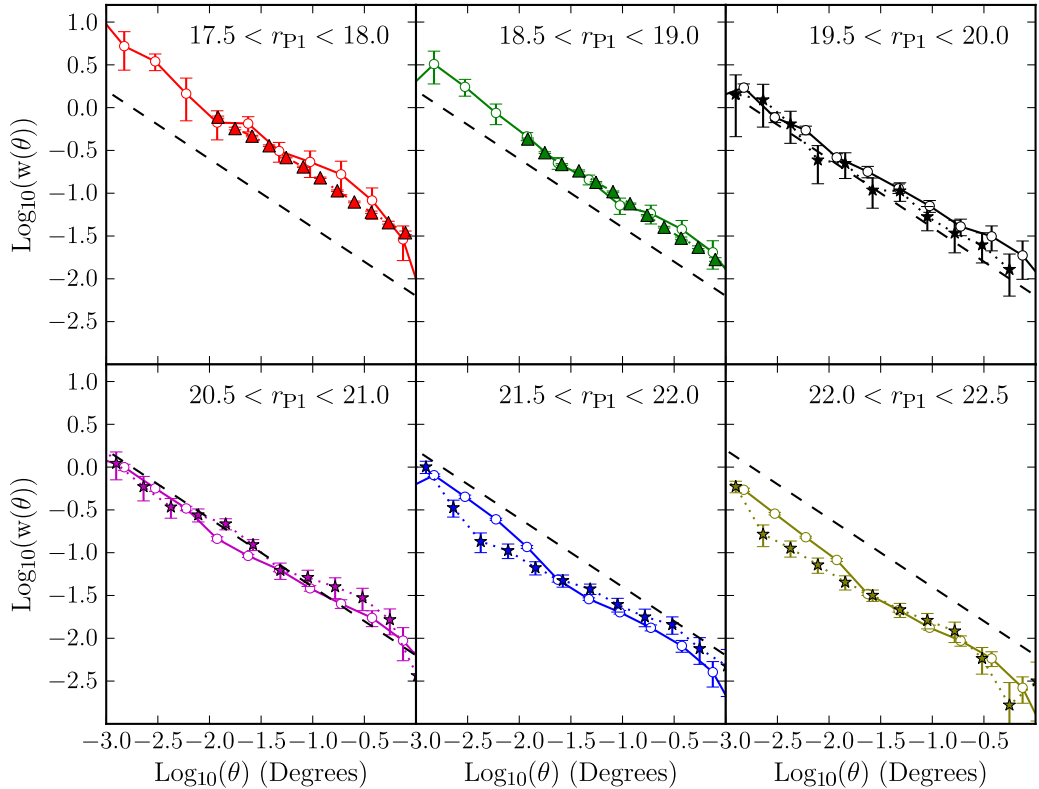}
 \caption{The lines and open points with error bars show the angular clustering of PS1 galaxies in the SAS2 region, for different magnitude ranges as indicated by the legend. Clustering measurements from Christodoulou et al (2012). for similar magnitude ranges from the full area of SDSS DR7 are plotted as triangles. The stars with error bars are measurements of clustering from the MD07 fields for Foucaud et al. (in preparation). Error bars on our measurements are estimated with 9 jack-knife re-samplings. The dashed line is a reference power law added to all of our clustering plots.}\label{fig:clustering}
\end{center}
\end{figure*}

In Fig.~\ref{fig:clustering} we show the angular correlation function measurements, using all the depth corrections described, down to $\rps=22.5$ in $0.5$ mag steps where we expect the clustering to still be reliable from Fig.~\ref{fig:clusteringSNRCuts}(a). The angular clustering results from the whole of SDSS DR7 measured for \cite{christo12}, and measurements of the clustering of fainter galaxies from PS1 MD07 from Foucaud et al. (in preparation) are also shown. Again, our clustering measurements and those of 
Foucaud et al. (in preparation) have been corrected for stellar contamination. 

The bright measurements are consistent within errors with the measurements made for \cite{christo12}. The fainter bins have power law shapes and lower amplitudes than the brighter bins, and agree with the MD07 measurements. Fig.~\ref{fig:clustering} is a positive indication that PS1, combined with these depth corrections, can measure clustering to fainter magnitudes than existing wide field optical surveys. 

\subsection{Clustering of Stars and False Positives}\label{sec:starFalse}
\begin{figure}
\begin{center}
 \includegraphics[width=3.6in]{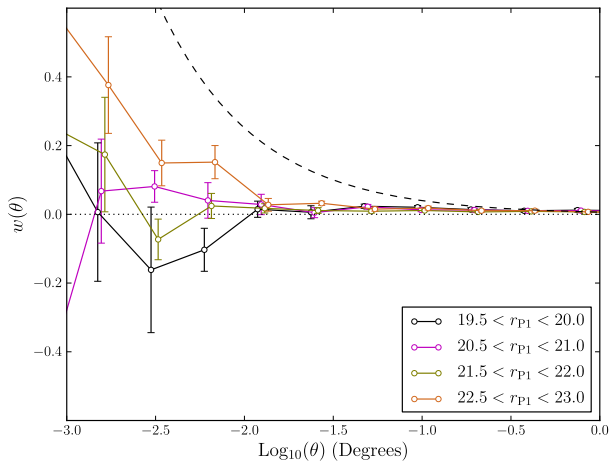}
 \caption{The angular correlation of objects not classed as galaxies by our adopted star and galaxy separator, split by magnitude, as indicated by the key. The dashed line is a reference power law added to all of our clustering plots, the dotted line marks no clustering.}\label{fig:starClustering}
\end{center}
\end{figure}

As it is expected that some contamination of our galaxy sample will occur due to stars and false positives, we estimate their effect on clustering by measuring their correlation functions. We begin by looking at stars; Fig.~\ref{fig:starClustering} gives the clustering of objects classified as stars by our separator. We do not correct these objects for extinction in this plot, as it is unclear that this would be appropriate. We have so far assumed that stars are distributed fairly uniformly across SAS2, and so simply affect the amplitude of the galaxy clustering. The brighter stellar bins do indeed show a less scale dependent signal than the galaxy samples, which is much weaker than the galaxy clustering except on the largest scales. 

\begin{figure}
\begin{center}
 \includegraphics[width=3.6in]{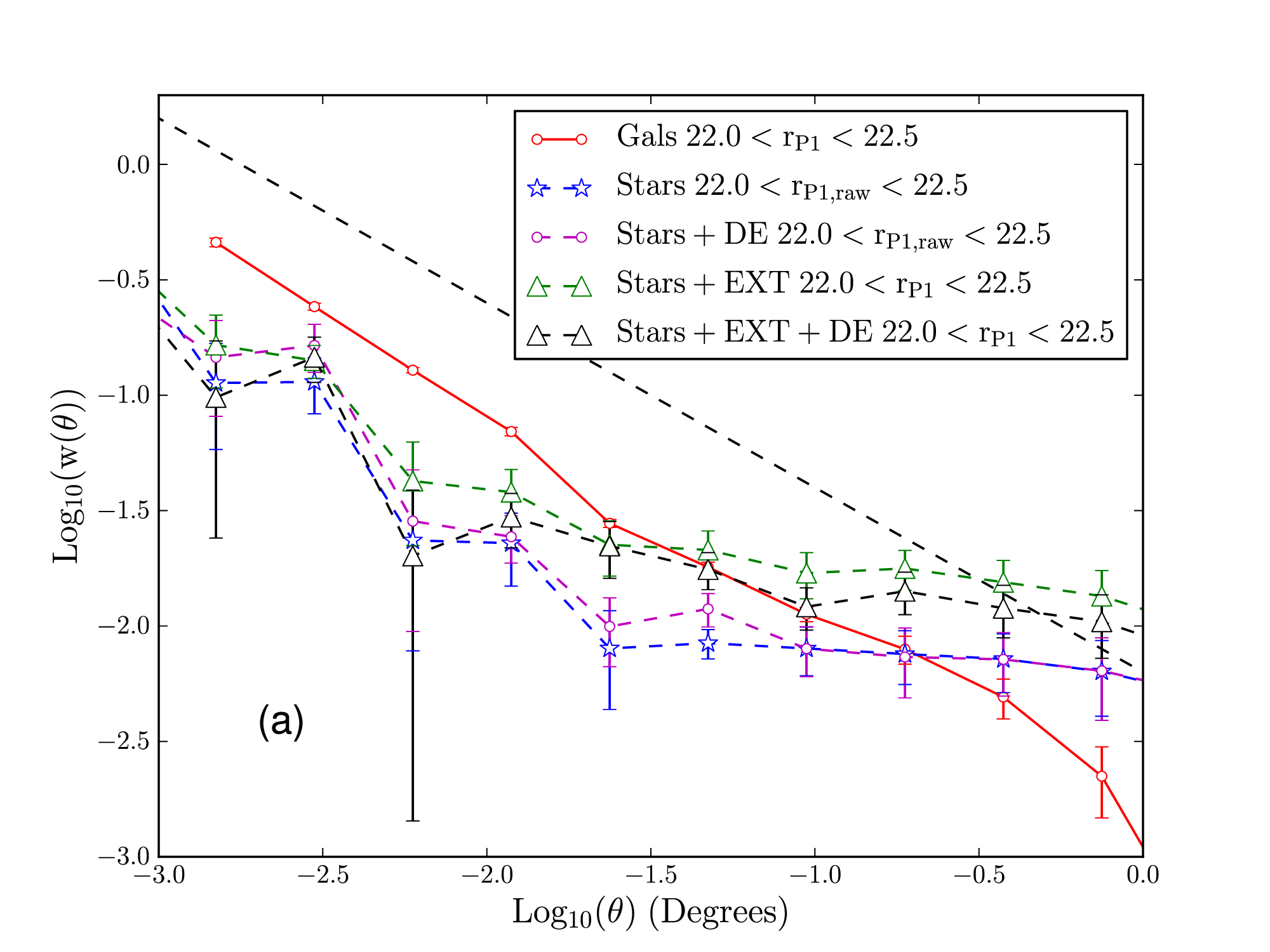}
 \includegraphics[width=3.6in]{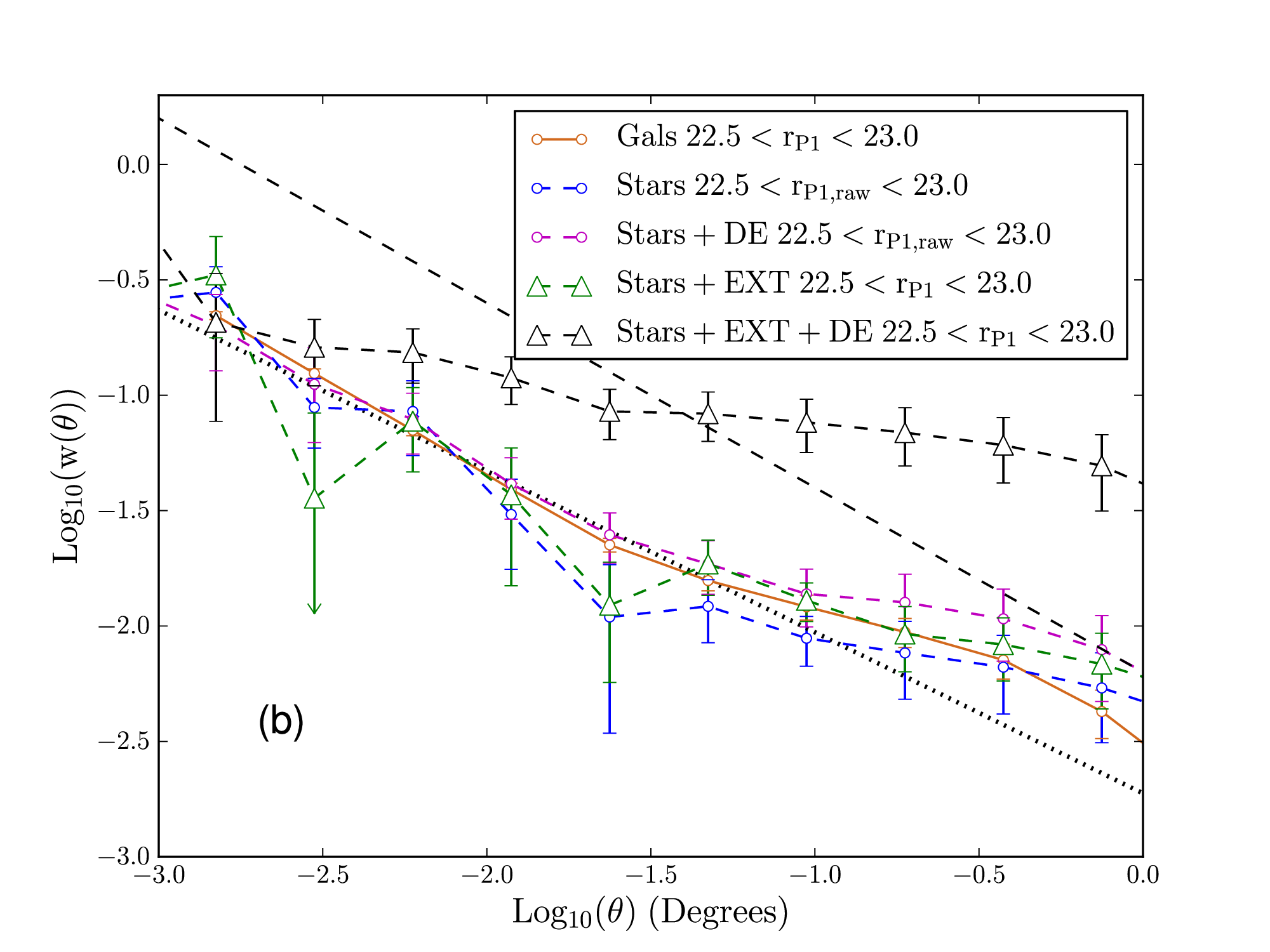}
 \caption{({\emph a}) Measurements of the angular correlation function of stars and galaxies. Dashed lines show the clustering of objects classed as stars, for measurements with either extinction corrections (EXT), detection efficiency corrections (DE) or both (DE+EXT) applied as indicated in the legend. The solid lines show the clustering of galaxies, with detection efficiency corrections and extinction corrections applied. ({\emph b}) As panel (a) but for a fainter magnitude range. The dotted gives the power law used to correct the clustering measurements in this panel for the integral constraint.}\label{fig:starClusteringlog1}
\end{center}
\end{figure}

Whilst we expect the clustering of stars to be weaker than that of the galaxies, we do not necessarily expect the stars to be unclustered. Stars appear in star clusters and 
gradients in stellar density exist due to the structure of the Milky Way. Measurements of the angular correlation function of stars have shown it to be flat and non-zero 
on larger scales \citep[e.g.][]{ross11, myers06}. In Fig~\ref{fig:starClusteringlog1}(a) we compare the clustering of faint galaxies to that of stars. We detect clustering in 
the stars which is weaker than the galaxies on small scales but stronger than the galaxies on larger scales. As such one could argue that the small scale clustering of stars 
is caused by contamination of the stellar sample by galaxies, whilst the large scale clustering of stars cannot be attributed to the galaxies. 
The clustering of stars in Fig~\ref{fig:starClusteringlog1}(a) is fairly insensitive to detection efficiency corrections. In contrast, extinction correcting the sample of 
stars enhances their clustering. This latter observation is concordant with the picture of stars having spatial density variations caused by the structure of the Milky Way. 
This is because one would expect dust to be correlated with the Milky Way's structure, and as such extinction correcting the stars would act to enhance spatial structure in 
the stellar sample. The enhancement of clustering signal after extinction correction is the opposite of what one would expect for galaxies. The flat angular correlation
function we measure on larger scales for stars is in accord with the shape reported in the literature for brighter stellar samples \citep[e.g.][]{ross11, myers06}.

In Fig~\ref{fig:starClusteringlog1}(b) we compare the clustering of a fainter sample of galaxies to that of stars. The clustering of stars in Fig~\ref{fig:starClusteringlog1}(b) 
has a similar amplitude to that of the the brighter stars; the galaxy clustering is weaker however, such that the stars and galaxies have a similar amplitude of clustering
on all scales. Again we do not believe the clustering in the star sample can be caused by galaxy contamination alone. This is because the clustering of galaxies in the 
stellar sample should be diluted by the stars in the sample and the resultant correlation function should have a lower amplitude than the galaxy sample. As in 
Fig~\ref{fig:starClusteringlog1}(a), extinction correcting the stellar sample boosts its clustering, though the effect is smaller than for the brighter magnitude bin. Detection 
efficiency corrections also boost the clustering of the stellar sample in Fig~\ref{fig:starClusteringlog1}(b). The likely cause of this is that the corrections are 
calibrated on galaxies which are harder to detect at these magnitudes, because they are extended. The combined effect of applying extinction corrections and 
detection efficiency corrections greatly boosts the clustering. This can be understood since their effect on the clustering will be compounded by the fact that 
extinction corrections bring objects with fainter observed magnitudes into the sample. These objects at fainter magnitudes have a 
larger detection efficiency correction and, since the detection efficiency corrections are based on galaxies, may artificially boost the stellar clustering further.     

Clearly the clustering of stars and the effect of stellar contamination on galaxy clustering measurements will have to be further studied. For the full 3$\pi$ survey 
measurements of the distribution of stars in the Milky Way could be used to attempt to model these effects. Cross correlating galaxy samples with stellar samples is also an 
important test we will carry out with the full 3$\pi$ data, which will allow us to further study the effects of misclassification and stellar contamination. For this work, 
the clustering of stars and contamination of the galaxy sample could be boosting the estimates of the galaxy correlation function on large scales. This will be a larger 
effect for the fainter galaxy samples where the large scale clustering of stars has a higher amplitude than that of the galaxies and the stellar contamination fractions are 
larger. An expression relating the true angular correlation function of galaxies, $w_{\rm{gg}}$, to the measured correlation function, $w_{\rm{measured}}$, given the angular 
correlation function of stars, $w_{\rm{ss}}$, can be found in \citet{myers06},
\begin{equation}
\label{eq:starCon}
w_{\rm{measured}}(\theta) = (1 - f)^2w_{\rm{gg}} + f^2w_{\rm{ss}} -\epsilon(\theta)
\end{equation}
where $\epsilon(\theta)$ is a very small cross term which is expected to be too small to influence our results. Eq.~\ref{eq:starCon} was derived by \citet{myers06} for the
\citet{landySzalay} estimator, but the \citet{landySzalay} gives very similar (much smaller than the error bars) results to the \citet{hamilton} estimator for our samples
so we can still use Eq. \ref{eq:starCon} to estimate the effect of stellar contamination. On small scales where $w_{\rm{ss}} \ll w_{\rm{gg}}$, Eq.~\ref{eq:starCon} reduces to
the $(1 - f)^2$ amplitude scaling we have used thus far. On larger scales and for fainter galaxy samples the star clustering can be stronger than the galaxy clustering. Using
the measured clustering of the extinction and detection efficiency corrected star samples we can estimate the effect of stars on the galaxy clustering. We do this by 
correcting the measured galaxy clustering using Eq.~\ref{eq:starCon} and comparing the result to using the simple $(1-f)^2$ correction we adopted. The faintest bin, 
$22.5<\rps<23.0$,  has a contamination fraction of $f=0.1$ (from Fig.~\ref{fig:sgCompl}) which leads to an enhancement of the galaxy clustering signal by clustered, 
stellar contaminants of $30\%$ and $18\%$ at 0.7 and 0.3 degrees respectively. For the brighter magnitude bin of $22.0<\rps<22.5$, with $f=0.08$, this drops to  
$6\%$ and $3\%$ for 0.7 and 0.3 degrees respectively. All of these differences are smaller than the errorbars, and as such adopting the $(1-f)^2$ for this data, 
instead of a more thorough modelling of stellar contamination, does not effect our conclusions. For the full 3$\pi$ survey where clustering on larger spatial 
scales will be measured this issue will have to be revisited. 

\begin{figure}
\begin{center}
 \includegraphics[width=3.6in]{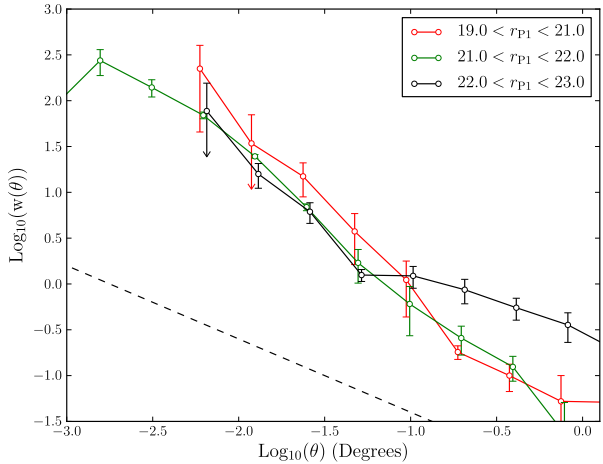}
 \caption{The angular correlation of objects cut by the extreme Kron minus PSF magnitude threshold given in Table \ref{table:coeff}, split by magnitude (see key). These objects are mostly false positives. Bins where one or more of the jack-knife regions have undefined clustering measurements, due to zero data-random or random-random pairs at that separation, have been omitted. The dashed line is a reference power law added to all of our clustering plots.} \label{fig:falseClustering}
\end{center}
\end{figure}

Fig.~\ref{fig:falseClustering} gives the clustering of objects with extreme $\Delta_{\rm{kron-psf}}$, removed by our cut in Fig.~\ref{fig:sgHist}, split into three magnitude bins. These objects are thought to be false positives. We see that, unfortunately, these objects have a strong clustering signal. This signal is well described by a power law that is steeper than the galaxy correlation functions. This is as false positives tend to appear in clumps around image artifacts and around real objects (see Paper I). Fortunately for magnitude bins brighter than $\rps=21.0$ false positives make up less than $1\%$ of the data (Fig.~\ref{fig:hist}) and are likely to have a negligable effect on clustering. For the fainter bins shown here, $\rps>21.0$, clustering could be affected by the false positives which can be as prevalent as $8-10\%$ of the sources. Remember that 
Fig.~\ref{fig:falseClustering} is measured from objects removed by our cut, false positives that evade this cut and so contaminate the galaxy sample could have different 
clustering. As we do not know if the false positives which evade our extreme $\Delta_{\rm{kron-psf}}$ have the same clustering as the objects in 
Fig.~\ref{fig:falseClustering}, Eq.~\ref{eq:starCon} cannot be used to estimate their effect on clustering. 

Improvements in the modelling of image artifacts will help ameliorate the problem of clustered false positives. Additionally requiring detections in multiple bands can 
also be effective in eliminating false positives.

\section{Discussion and Conclusions}\label{sec:diss}
We have presented methods of star and galaxy separation, angular masking and completeness corrections for PS1. Our star and galaxy separation approach uses fake images to
identify cuts in  $\Delta_{\rm{kron-psf}}$ that yield galaxy samples. Our separator is 92\%-98\% complete with less than around $10\%$ stellar contamination down to a 
magnitude of $\rps<23.0$. However, SAS2 has uniform properties so before applying this to the full 3$\pi$ data we need to test and calibrate the star/galaxy separator for different seeing and background noise. It is likely that the galaxy distribution in $\Delta_{\rm{kron-psf}}$ will depend on seeing. Changing the PSF of an image has a different effect on the surface brightness of stars and galaxies and this will drive a change in a galaxy's measured PSF magnitude. Ultimately a more sophisticated star and galaxy separator with better completeness and less contamination 
will need to be developed. Using the colours of galaxies \citep[e.g][]{saglia12} and other morphological measurements, such as galaxy size, are promising 
avenues to achieve this with PS1 data.

We present a method of generating angular masks for PS1 3$\pi$ data, using a statistical approach to define the size of masked regions around bright stars. The relation between mask size and magnitude may vary across the much larger 3$\pi$ field and as such the relation may need to be re-calibrated on the full data. 
We also presented our binned-up variance maps, which we have used to develop a method of correcting PS1 measurements for spatially varying depth. A question left to address is what binning scale to choose for masks and maps of the whole survey. One has to balance accuracy with the computational costs of using large amounts of data. Ultimately the mask size will also depend on the science goals; BAO measurements for example will be less sensitive to small scale systematics than galaxy formation studies using small scale clustering.

Some further questions related to our depth corrections will have to be addressed in future work. Firstly, we need to test how well our $\rm{SNR}$ technique applies across a larger field with more variable PSFs and depths. One way to calibrate and test our method for the full survey is to utilise the 10 Medium Deep fields, which are scattered across the sky.  Using surveys in addition to Stripe 82, such as the Medium Deep surveys, can also help remove the effects of false positives from our measurements of the probability of detection versus fiducial SNR. Additionally our assumption that all galaxies have the same detection efficiency properties will have to be further 
explored, perhaps by studying clustering as a function of colour. Our comparisons of detection efficiency for red and blue Stripe 82 galaxies are a positive indication that this is a valid assumption. We can also gain more insight into our depth correction method by utilising our synthetic images to simulate more greatly varying PSFs and backgrounds.  

One important test of our method is excluding regions which fail to meet some SNR requirement and testing if clustering measurements from them agree with data with a less conservative cut. This test was demonstrated in Fig.~\ref{fig:clusteringSNRCuts} for SAS2 data but will have to be applied to the full 3$\pi$ data. The application of this test to the the full 3$\pi$ data may be more fruitful as the much larger area will decrease the random errors on the measurement and make any systematics more apparent. Ultimately this SNR cut can be used as a free parameter in our method, which can be varied to ensure science results are not sensitive to its value. It is also important when using this
method to choose magnitude limits which ensure the detected galaxies have clustering representive of the full galaxy sample. This can be achieved by using narrow magnitude
bins or by only applying this correction to fairly complete samples.

By applying our methods to a set of science verification data we show that in the PS1 3$\pi$ measurements of clustering show reasonable agreement with literature data down 
to a magnitude of $\rps<23.0$. Though tests using regions with different fiducial SNR limits, and the large size of correction needed at these faint magnitudes, 
suggest perhaps a limit of $\rps<22.5$ is a more reliable estimate of how faint we can use this method. These limits may change as the PS1 survey matures. At bright magnitudes we show agreement with the 
published angular correlation function estimates of \cite{christo12} and \cite{wang13}, fainter than this our measurements show the decrease in amplitude expected. Our 
fainter measurements agree, within error, with the measurements of \cite{hudon96}, \cite{roche99} and with Foucaud et al. (in preparation) for the threshold sample 
$\rps<23.0$. Our magnitude bin samples also agree within error with Foucaud et al. (in preparation) down to a limit of $\rps < 22.5$. We also demonstrate our method yields 
sensible measurements of the number counts of galaxies, with $\rps$-band counts, showing agreement with published data.

One difficulty with the literature comparisons is the relative deficit of faint $r$-band comparison data, especially from fields large enough to test the scales where our 
correction is strongest. In places large sample variance could be masking residual systematics caused by the spatially varying depth. Future work will be 
able to further test our depth correction technique in several ways. Firstly, the extension of this work to different bands will allow a larger number of literature 
comparisons to be made. Additionally, combining the data across multiple bands will allow us to test the depth correction technique with more complex selection criteria, 
such as colour. Finally, using the full 3$\pi$ data will greatly decrease the random errors in the SAS2 measurements, making systematics more apparent.

Clustered false positives are a potential limitation to measuring clustering, but these only affect the fainter magnitude bins and this problem should be improved by future efforts in understanding the instrumental signature of the PS1 camera. Additionally, matching between bands, which will be necessary for photometric redshifts, will go a long way in removing these false positives as image defects are very unlikely to be located in the same place in multiple bands.

Further issues not covered in this work, but which will still have to be considered when utilising the full survey also include how extinction corrections and stellar contamination affect the measured clustering signal. Issues such as these are common to many large galaxy surveys and there are approaches in the literature to deal with them \citep[e.g.][]{ross12, wang13}.     

Once the 3$\pi$ survey is complete we will apply these methods to the full survey, which is due to be completed by around January 2014, with data reduction complete by mid 2014 (Magnier et al. in preparation).  If the techniques developed here are successfully applied, the PS1 $3\pi$ survey will be able to push forward our understanding of cosmology and galaxy formation. One particularly exciting application will be to measure the Integrated Sachs-Wolfe effect by cross correlating PS1 galaxies with CMB data. The large area of $3\pi$ will be ideal for minimizing sample variance and false positives will become less of an issue as they are not correlated with the CMB. 

\section*{Acknowledgements}

DJF acknowledges the support of an STFC studentship (ST/F007299/1). The authors acknowledge a grant with the RCUK reference ST/F001166/1. PN acknowledges the support of the Royal Society through the award of a University Research Fellowship and the European Research Council, through receipt of a Starting Grant (DEGAS-259586). PD acknowledges support from an ERC Starting Grant (DEGAS-259586). We also thank the referee for their comments on this paper. 

The Pan-STARRS1 Surveys (PS1) have been made possible through contributions of the Institute for Astronomy, the University of Hawaii, the Pan-STARRS Project Office, the Max-Planck Society and its participating institutes, the Max Planck Institute for Astronomy, Heidelberg and the Max Planck Institute for Extraterrestrial Physics, Garching, The Johns Hopkins University, Durham University, the University of Edinburgh, Queen's University Belfast, the Harvard-Smithsonian Center for Astrophysics, the Las Cumbres Observatory Global Telescope Network Incorporated, the National Central University of Taiwan, the Space Telescope Science Institute, the National Aeronautics and Space Administration under Grant No. NNX08AR22G issued through the Planetary Science Division of the NASA Science Mission Directorate, the National Science Foundation under Grant No. AST-1238877, and the University of Maryland.

The clustering measurements for this paper where carried out on the GPUs of the Cosmology Machine supercomputer at the Institute for Computational Cosmology, Durham. The Cosmology Machine is part of the DiRAC Facility jointly funded by the STFC, the Large Facilities and Capital Fund of BIS, and Durham University.

Durham University's membership of PS1 was made possible through the generous support of the Ogden Trust.

The PS1 catalogue data is currently restricted to the private PS1 consortium but will be made available through the Space Telescope Science Institute when the proprietary 
period is complete. Access to the measurements published here can be obtained from the lead author.

\bibliographystyle{mn2e}
\bibliography{mybibliography}

\begin{thebibliography}{53}
\expandafter\ifx\csname natexlab\endcsname\relax\def\natexlab#1{#1}\fi

\bibitem[{{Annis} {et~al}\mbox{.}(2011){Annis}, {Soares-Santos}, {Strauss},
  {Becker}, {Dodelson}, {Fan}, {Gunn}, {Hao}, {Ivezic}, {Jester}, {Jiang},
  {Johnston}, {Kubo}, {Lampeitl}, {Lin}, {Lupton}, {Miknaitis}, {Seo}, {Simet},
  \& {Yanny}}]{annis11}
{Annis} J. {et~al.}, 2011, ArXiv e-prints

\bibitem[{Bard {et~al}\mbox{.}(2012)Bard, Bellis, Allen, Yepremyan, \&
  Kratochvil}]{bard}
Bard D., Bellis M., Allen M.~T., Yepremyan H., Kratochvil J.~M., 2012

\bibitem[{{Bertin} \& {Arnouts}(1996)}]{bertin}
{Bertin} E., {Arnouts} S., 1996, \aaps, 117, 393

\bibitem[{{Blanton} {et~al}\mbox{.}(2001){Blanton}, {Dalcanton}, {Eisenstein},
  {Loveday}, {Strauss}, {SubbaRao}, {Weinberg}, {Anderson}, {Annis}, {Bahcall},
  {Bernardi}, {Brinkmann}, {Brunner}, {Burles}, {Carey}, {Castander},
  {Connolly}, {Csabai}, {Doi}, {Finkbeiner}, {Friedman}, {Frieman}, {Fukugita},
  {Gunn}, {Hennessy}, {Hindsley}, {Hogg}, {Ichikawa}, {Ivezi{\'c}}, {Kent},
  {Knapp}, {Lamb}, {Leger}, {Long}, {Lupton}, {McKay}, {Meiksin}, {Merelli},
  {Munn}, {Narayanan}, {Newcomb}, {Nichol}, {Okamura}, {Owen}, {Pier}, {Pope},
  {Postman}, {Quinn}, {Rockosi}, {Schlegel}, {Schneider}, {Shimasaku},
  {Siegmund}, {Smee}, {Snir}, {Stoughton}, {Stubbs}, {Szalay}, {Szokoly},
  {Thakar}, {Tremonti}, {Tucker}, {Uomoto}, {Vanden Berk}, {Vogeley},
  {Waddell}, {Yanny}, {Yasuda}, \& {York}}]{blanton2001}
{Blanton} M.~R. {et~al.}, 2001, \aj, 121, 2358

\bibitem[{{Bottini} {et~al}\mbox{.}(2005){Bottini}, {Garilli}, {Maccagni},
  {Tresse}, {Le Brun}, {Le F{\`e}vre}, {Picat}, {Scaramella}, {Scodeggio},
  {Vettolani}, {Zanichelli}, {Adami}, {Arnaboldi}, {Arnouts}, {Bardelli},
  {Bolzonella}, {Cappi}, {Charlot}, {Ciliegi}, {Contini}, {Foucaud},
  {Franzetti}, {Guzzo}, {Ilbert}, {Iovino}, {McCracken}, {Marano}, {Marinoni},
  {Mathez}, {Mazure}, {Meneux}, {Merighi}, {Paltani}, {Pollo}, {Pozzetti},
  {Radovich}, {Zamorani}, \& {Zucca}}]{bottini05}
{Bottini} D. {et~al.}, 2005, \pasp, 117, 996

\bibitem[{{Bower} {et~al}\mbox{.}(2006){Bower}, {Benson}, {Malbon}, {Helly},
  {Frenk}, {Baugh}, {Cole}, \& {Lacey}}]{bower06}
{Bower} R.~G., {Benson} A.~J., {Malbon} R., {Helly} J.~C., {Frenk} C.~S.,
  {Baugh} C.~M., {Cole} S., {Lacey} C.~G., 2006, \mnras, 370, 645

\bibitem[{{Christodoulou} {et~al}\mbox{.}(2012){Christodoulou}, {Eminian}, \&
  {Loveday}}]{christo12}
{Christodoulou} L., {Eminian} C., {Loveday} J., 2012, \mnras, 425, 1527

\bibitem[{{de Vaucouleurs}(1948)}]{deVaucouleurs}
{de Vaucouleurs} G., 1948, Annales d'Astrophysique, 11, 247

\bibitem[{{Dutton} {et~al}\mbox{.}(2011){Dutton}, {van den Bosch}, {Faber},
  {Simard}, {Kassin}, {Koo}, {Bundy}, {Huang}, {Weiner}, {Cooper}, {Newman},
  {Mozena}, \& {Koekemoer}}]{dutton2011}
{Dutton} A.~A. {et~al.}, 2011, \mnras, 410, 1660

\bibitem[{{Ferrarese} {et~al}\mbox{.}(2006){Ferrarese}, {C{\^o}t{\'e}},
  {Jord{\'a}n}, {Peng}, {Blakeslee}, {Piatek}, {Mei}, {Merritt},
  {Milosavljevi{\'c}}, {Tonry}, \& {West}}]{Ferrarese2006}
{Ferrarese} L. {et~al.}, 2006, \apjs, 164, 334

\bibitem[{{Finlator} {et~al}\mbox{.}(2000){Finlator}, {Ivezi{\'c}}, {Fan},
  {Strauss}, {Knapp}, {Lupton}, {Gunn}, {Rockosi}, {Anderson}, {Csabai},
  {Hennessy}, {Hindsley}, {McKay}, {Nichol}, {Schneider}, {Smith}, {York}, \&
  {SDSS Collaboration}}]{finlator}
{Finlator} K. {et~al.}, 2000, \aj, 120, 2615

\bibitem[{Frigo \& Johnson(2005)}]{FFTW05}
Frigo M., Johnson S.~G., 2005, Proceedings of the IEEE, 93, 216, special issue
  on ``Program Generation, Optimization, and Platform Adaptation''

\bibitem[{{Graham} \& {Driver}(2005)}]{Graham05}
{Graham} A.~W., {Driver} S.~P., 2005, Publications of the Astronomical Society
  of Australia, 22, 118

\bibitem[{{Hamilton}(1993)}]{hamilton}
{Hamilton} A.~J.~S., 1993, \apj, 417, 19

\bibitem[{{Hodapp} {et~al}\mbox{.}(2004){Hodapp}, {Kaiser}, {Aussel},
  {Burgett}, {Chambers}, {Chun}, {Dombeck}, {Douglas}, {Hafner}, {Heasley},
  {Hoblitt}, {Hude}, {Isani}, {Jedicke}, {Jewitt}, {Laux}, {Luppino}, {Lupton},
  {Maberry}, {Magnier}, {Mannery}, {Monet}, {Morgan}, {Onaka}, {Price}, {Ryan},
  {Siegmund}, {Szapudi}, {Tonry}, {Wainscoat}, \& {Waterson}}]{hodapp}
{Hodapp} K.~W. {et~al.}, 2004, Astronomische Nachrichten, 325, 636

\bibitem[{{Huang} {et~al}\mbox{.}(2001){Huang}, {Thompson}, {K{\"u}mmel},
  {Meisenheimer}, {Wolf}, {Beckwith}, {Fockenbrock}, {Fried}, {Hippelein}, {von
  Kuhlmann}, {Phleps}, {R{\"o}ser}, \& {Thommes}}]{huang01}
{Huang} J.-S. {et~al.}, 2001, \aap, 368, 787

\bibitem[{{Hudon} \& {Lilly}(1996)}]{hudon96}
{Hudon} J.~D., {Lilly} S.~J., 1996, \apj, 469, 519

\bibitem[{{Ilbert} {et~al}\mbox{.}(2005){Ilbert}, {Tresse}, {Zucca},
  {Bardelli}, {Arnouts}, {Zamorani}, {Pozzetti}, {Bottini}, {Garilli}, {Le
  Brun}, {Le F{\`e}vre}, {Maccagni}, {Picat}, {Scaramella}, {Scodeggio},
  {Vettolani}, {Zanichelli}, {Adami}, {Arnaboldi}, {Bolzonella}, {Cappi},
  {Charlot}, {Contini}, {Foucaud}, {Franzetti}, {Gavignaud}, {Guzzo}, {Iovino},
  {McCracken}, {Marano}, {Marinoni}, {Mathez}, {Mazure}, {Meneux}, {Merighi},
  {Paltani}, {Pello}, {Pollo}, {Radovich}, {Bondi}, {Bongiorno}, {Busarello},
  {Ciliegi}, {Lamareille}, {Mellier}, {Merluzzi}, {Ripepi}, \&
  {Rizzo}}]{ilbert05}
{Ilbert} O. {et~al.}, 2005, \aap, 439, 863

\bibitem[{{Jian} {et~al}\mbox{.}(2013){Jian}, {Lin}, {Chiueh}, {Lin}, {Liu},
  {Merson}, {Baugh}, {Huang}, {Chen}, {Foucaud}, {Murphy}, {Cole}, {Burgett},
  \& {Kaiser}}]{jian2013}
{Jian} H.-Y. {et~al.}, 2013, ArXiv e-prints

\bibitem[{{Kaiser} {et~al}\mbox{.}(2002){Kaiser}, {Aussel}, {Burke},
  {Boesgaard}, {Chambers}, {Chun}, {Heasley}, {Hodapp}, {Hunt}, {Jedicke},
  {Jewitt}, {Kudritzki}, {Luppino}, {Maberry}, {Magnier}, {Monet}, {Onaka},
  {Pickles}, {Rhoads}, {Simon}, {Szalay}, {Szapudi}, {Tholen}, {Tonry},
  {Waterson}, \& {Wick}}]{kaiser02}
{Kaiser} N. {et~al.}, 2002, in Society of Photo-Optical Instrumentation
  Engineers (SPIE) Conference Series, Vol. 4836, Society of Photo-Optical
  Instrumentation Engineers (SPIE) Conference Series, {Tyson} J.~A., {Wolff}
  S., eds., pp. 154--164

\bibitem[{{Kashikawa} {et~al}\mbox{.}(2004){Kashikawa}, {Shimasaku}, {Yasuda},
  {Ajiki}, {Akiyama}, {Ando}, {Aoki}, {Doi}, {Fujita}, {Furusawa}, {Hayashino},
  {Iwamuro}, {Iye}, {Karoji}, {Kobayashi}, {Kodaira}, {Kodama}, {Komiyama},
  {Matsuda}, {Miyazaki}, {Mizumoto}, {Morokuma}, {Motohara}, {Murayama},
  {Nagao}, {Nariai}, {Ohta}, {Okamura}, {Ouchi}, {Sasaki}, {Sato}, {Sekiguchi},
  {Shioya}, {Tamura}, {Taniguchi}, {Umemura}, {Yamada}, \&
  {Yoshida}}]{kashikawa04}
{Kashikawa} N. {et~al.}, 2004, \pasj, 56, 1011

\bibitem[{{Kron}(1980)}]{1980ApJS...43..305K}
{Kron} R.~G., 1980, \apjs, 43, 305

\bibitem[{{Landy} \& {Szalay}(1993)}]{landySzalay}
{Landy} S.~D., {Szalay} A.~S., 1993, \apj, 412, 64

\bibitem[{{Le F{\`e}vre} {et~al}\mbox{.}(2005){Le F{\`e}vre}, {Vettolani},
  {Garilli}, {Tresse}, {Bottini}, {Le Brun}, {Maccagni}, {Picat}, {Scaramella},
  {Scodeggio}, {Zanichelli}, {Adami}, {Arnaboldi}, {Arnouts}, {Bardelli},
  {Bolzonella}, {Cappi}, {Charlot}, {Ciliegi}, {Contini}, {Foucaud},
  {Franzetti}, {Gavignaud}, {Guzzo}, {Ilbert}, {Iovino}, {McCracken}, {Marano},
  {Marinoni}, {Mathez}, {Mazure}, {Meneux}, {Merighi}, {Paltani}, {Pell{\`o}},
  {Pollo}, {Pozzetti}, {Radovich}, {Zamorani}, {Zucca}, {Bondi}, {Bongiorno},
  {Busarello}, {Lamareille}, {Mellier}, {Merluzzi}, {Ripepi}, \&
  {Rizzo}}]{fevre05}
{Le F{\`e}vre} O. {et~al.}, 2005, \aap, 439, 845

\bibitem[{{Lupton} {et~al}\mbox{.}(1999){Lupton}, {Gunn}, \& {Szalay}}]{lupton}
{Lupton} R.~H., {Gunn} J.~E., {Szalay} A.~S., 1999, \aj, 118, 1406

\bibitem[{{Magnier}(2006)}]{magnier2006}
{Magnier} E., 2006, in The Advanced Maui Optical and Space Surveillance
  Technologies Conference

\bibitem[{{Magnier} {et~al}\mbox{.}(2013){Magnier}, {Schlafly}, {Finkbeiner},
  {Juric}, {Tonry}, {Burgett}, {Chambers}, {Flewelling}, {Kaiser}, {Kudritzki},
  {Morgan}, {Price}, {Sweeney}, \& {Stubbs}}]{magnier13}
{Magnier} E.~A. {et~al.}, 2013, \apjs, 205, 20

\bibitem[{{Martin} {et~al}\mbox{.}(2013){Martin}, {Slater}, {Schlafly},
  {Morganson}, {Rix}, {Bell}, {Laevens}, {Bernard}, {Ferguson}, {Finkbeiner},
  {Burgett}, {Chambers}, {Hodapp}, {Kaiser}, {Kudritzki}, {Magnier}, {Morgan},
  {Price}, {Tonry}, \& {Wainscoat}}]{martin2013}
{Martin} N.~F. {et~al.}, 2013, \apj, 772, 15

\bibitem[{{McCracken} {et~al}\mbox{.}(2003){McCracken}, {Radovich}, {Bertin},
  {Mellier}, {Dantel-Fort}, {Le F{\`e}vre}, {Cuillandre}, {Gwyn}, {Foucaud}, \&
  {Zamorani}}]{mccracken03}
{McCracken} H.~J. {et~al.}, 2003, \aap, 410, 17

\bibitem[{{Merson} {et~al}\mbox{.}(2013){Merson}, {Baugh}, {Helly},
  {Gonzalez-Perez}, {Cole}, {Bielby}, {Norberg}, {Frenk}, {Benson}, {Bower},
  {Lacey}, \& {Lagos}}]{merson}
{Merson} A.~I. {et~al.}, 2013, \mnras, 429, 556

\bibitem[{{Metcalfe} {et~al}\mbox{.}(2013){Metcalfe}, {Farrow}, {Cole},
  {Draper}, {Norberg}, {Burgett}, {Chambers}, {Denneau}, {Flewelling},
  {Kaiser}, {Kudritzki}, {Magnier}, {Morgan}, {Price}, {Sweeney}, {Tonry},
  {Wainscoat}, \& {Waters}}]{metcalfe2013}
{Metcalfe} N. {et~al.}, 2013, \mnras, 435, 1825

\bibitem[{{Mo} {et~al}\mbox{.}(2010){Mo}, {van den Bosch}, \& {White}}]{mo}
{Mo} H., {van den Bosch} F.~C., {White} S., 2010, {Galaxy Formation and
  Evolution}

\bibitem[{{Myers} {et~al}\mbox{.}(2006){Myers}, {Brunner}, {Richards},
  {Nichol}, {Schneider}, {Vanden Berk}, {Scranton}, {Gray}, \&
  {Brinkmann}}]{myers06}
{Myers} A.~D. {et~al.}, 2006, \apj, 638, 622

\bibitem[{{Padilla} \& {Strauss}(2008)}]{padilla08}
{Padilla} N.~D., {Strauss} M.~A., 2008, \mnras, 388, 1321

\bibitem[{{Petrosian}(1976)}]{petrosian1976}
{Petrosian} V., 1976, \apjl, 209, L1

\bibitem[{{Roche} \& {Eales}(1999)}]{roche99}
{Roche} N., {Eales} S.~A., 1999, \mnras, 307, 703

\bibitem[{{Ross} {et~al}\mbox{.}(2011){Ross}, {Ho}, {Cuesta}, {Tojeiro},
  {Percival}, {Wake}, {Masters}, {Nichol}, {Myers}, {de Simoni}, {Seo},
  {Hern{\'a}ndez-Monteagudo}, {Crittenden}, {Blanton}, {Brinkmann}, {da Costa},
  {Guo}, {Kazin}, {Maia}, {Maraston}, {Padmanabhan}, {Prada}, {Ramos},
  {Sanchez}, {Schlafly}, {Schlegel}, {Schneider}, {Skibba}, {Thomas}, {Weaver},
  {White}, \& {Zehavi}}]{ross11}
{Ross} A.~J. {et~al.}, 2011, \mnras, 417, 1350

\bibitem[{{Ross} {et~al}\mbox{.}(2012){Ross}, {Percival}, {S{\'a}nchez},
  {Samushia}, {Ho}, {Kazin}, {Manera}, {Reid}, {White}, {Tojeiro}, {McBride},
  {Xu}, {Wake}, {Strauss}, {Montesano}, {Swanson}, {Bailey}, {Bolton}, {Dorta},
  {Eisenstein}, {Guo}, {Hamilton}, {Nichol}, {Padmanabhan}, {Prada},
  {Schlegel}, {Maga{\~n}a}, {Zehavi}, {Blanton}, {Bizyaev}, {Brewington},
  {Cuesta}, {Malanushenko}, {Malanushenko}, {Oravetz}, {Parejko}, {Pan},
  {Schneider}, {Shelden}, {Simmons}, {Snedden}, \& {Zhao}}]{ross12}
{Ross} A.~J. {et~al.}, 2012, \mnras, 424, 564

\bibitem[{{Saglia} {et~al}\mbox{.}(2012){Saglia}, {Tonry}, {Bender}, {Greisel},
  {Seitz}, {Senger}, {Snigula}, {Phleps}, {Wilman}, {Bailer-Jones}, {Klement},
  {Rix}, {Smith}, {Green}, {Burgett}, {Chambers}, {Heasley}, {Kaiser},
  {Magnier}, {Morgan}, {Price}, {Stubbs}, \& {Wainscoat}}]{saglia12}
{Saglia} R.~P. {et~al.}, 2012, \apj, 746, 128

\bibitem[{{Schlafly} \& {Finkbeiner}(2011)}]{sch2011}
{Schlafly} E.~F., {Finkbeiner} D.~P., 2011, \apj, 737, 103

\bibitem[{{Schlafly} {et~al}\mbox{.}(2012){Schlafly}, {Finkbeiner},
  {Juri{\'c}}, {Magnier}, {Burgett}, {Chambers}, {Grav}, {Hodapp}, {Kaiser},
  {Kudritzki}, {Martin}, {Morgan}, {Price}, {Rix}, {Stubbs}, {Tonry}, \&
  {Wainscoat}}]{eddie2}
{Schlafly} E.~F. {et~al.}, 2012, \apj, 756, 158

\bibitem[{{Schlegel} {et~al}\mbox{.}(1998){Schlegel}, {Finkbeiner}, \&
  {Davis}}]{schlegel}
{Schlegel} D.~J., {Finkbeiner} D.~P., {Davis} M., 1998, \apj, 500, 525

\bibitem[{{S{\'e}rsic}(1963)}]{sersic}
{S{\'e}rsic} J.~L., 1963, Boletin de la Asociacion Argentina de Astronomia La
  Plata Argentina, 6, 41

\bibitem[{{Shen} {et~al}\mbox{.}(2003){Shen}, {Mo}, {White}, {Blanton},
  {Kauffmann}, {Voges}, {Brinkmann}, \& {Csabai}}]{shen}
{Shen} S., {Mo} H.~J., {White} S.~D.~M., {Blanton} M.~R., {Kauffmann} G.,
  {Voges} W., {Brinkmann} J., {Csabai} I., 2003, \mnras, 343, 978

\bibitem[{{Simard} {et~al}\mbox{.}(2011){Simard}, {Mendel}, {Patton},
  {Ellison}, \& {McConnachie}}]{simard}
{Simard} L., {Mendel} J.~T., {Patton} D.~R., {Ellison} S.~L., {McConnachie}
  A.~W., 2011, \apjs, 196, 11

\bibitem[{{Strauss} {et~al}\mbox{.}(2002){Strauss}, {Weinberg}, {Lupton},
  {Narayanan}, {Annis}, {Bernardi}, {Blanton}, {Burles}, {Connolly},
  {Dalcanton}, {Doi}, {Eisenstein}, {Frieman}, {Fukugita}, {Gunn},
  {Ivezi{\'c}}, {Kent}, {Kim}, {Knapp}, {Kron}, {Munn}, {Newberg}, {Nichol},
  {Okamura}, {Quinn}, {Richmond}, {Schlegel}, {Shimasaku}, {SubbaRao},
  {Szalay}, {Vanden Berk}, {Vogeley}, {Yanny}, {Yasuda}, {York}, \&
  {Zehavi}}]{strauss02}
{Strauss} M.~A. {et~al.}, 2002, \aj, 124, 1810

\bibitem[{{Tonry} {et~al}\mbox{.}(2008){Tonry}, {Burke}, {Isani}, {Onaka}, \&
  {Cooper}}]{tonry08}
{Tonry} J.~L., {Burke} B.~E., {Isani} S., {Onaka} P.~M., {Cooper} M.~J., 2008,
  in Society of Photo-Optical Instrumentation Engineers (SPIE) Conference
  Series, Vol. 7021, Society of Photo-Optical Instrumentation Engineers (SPIE)
  Conference Series

\bibitem[{{Tonry} {et~al}\mbox{.}(2012){Tonry}, {Stubbs}, {Lykke}, {Doherty},
  {Shivvers}, {Burgett}, {Chambers}, {Hodapp}, {Kaiser}, {Kudritzki},
  {Magnier}, {Morgan}, {Price}, \& {Wainscoat}}]{tonry12}
{Tonry} J.~L. {et~al.}, 2012, \apj, 750, 99

\bibitem[{{Valenti} {et~al}\mbox{.}(2010){Valenti}, {Smartt}, {Young}, {Smith},
  {Metcalfe}, {Bresolin}, {Kudritzki}, {Tonry}, {Price}, {Magnier}, {Chambers},
  {Kaiser}, {Morgan}, {Burgett}, {Heasley}, {Sweeney}, {Waters}, \&
  {Flewelling}}]{smartt}
{Valenti} S. {et~al.}, 2010, The Astronomer's Telegram, 2773, 1

\bibitem[{{Wang} {et~al}\mbox{.}(2013){Wang}, {Brunner}, \& {Dolence}}]{wang13}
{Wang} Y., {Brunner} R.~J., {Dolence} J.~C., 2013, \mnras, 432, 1961

\bibitem[{{Ward} {et~al}\mbox{.}(2011){Ward}, {Hutton}, {Mattila}, \&
  {Kotak}}]{ward}
{Ward} M.~J., {Hutton} S., {Mattila} S., {Kotak} R., 2011, in American
  Astronomical Society Meeting Abstracts \#218

\bibitem[{{Yasuda} {et~al}\mbox{.}(2001){Yasuda}, {Fukugita}, {Narayanan},
  {Lupton}, {Strateva}, {Strauss}, {Ivezi{\'c}}, {Kim}, {Hogg}, {Weinberg},
  {Shimasaku}, {Loveday}, {Annis}, {Bahcall}, {Blanton}, {Brinkmann},
  {Brunner}, {Connolly}, {Csabai}, {Doi}, {Hamabe}, {Ichikawa}, {Ichikawa},
  {Johnston}, {Knapp}, {Kunszt}, {Lamb}, {McKay}, {Munn}, {Nichol}, {Okamura},
  {Schneider}, {Szokoly}, {Vogeley}, {Watanabe}, \& {York}}]{yasuda2001}
{Yasuda} N. {et~al.}, 2001, \aj, 122, 1104

\bibitem[{{Zacharias} {et~al}\mbox{.}(2013){Zacharias}, {Finch}, {Girard},
  {Henden}, {Bartlett}, {Monet}, \& {Zacharias}}]{z13}
{Zacharias} N., {Finch} C.~T., {Girard} T.~M., {Henden} A., {Bartlett} J.~L.,
  {Monet} D.~G., {Zacharias} M.~I., 2013, \aj, 145, 44

\end{thebibliography}
\label{lastpage}

\end{document}